\documentclass[sigconf, nonacm]{acmart}
\usepackage{xcolor}
\usepackage{graphicx}
\usepackage{pifont}
\usepackage{enumitem}
\usepackage{subcaption}
\newcommand{\added}[1]{\textcolor{blue}{#1}}
\renewcommand{\added}[1]{#1}
\AtBeginDocument{%
  }

\acmConference[CHI '26]{Proceedings of the 2026 CHI Conference on Human Factors in Computing Systems}{April 13--17, 2026}{Barcelona, Spain}
\acmBooktitle{Proceedings of the 2026 CHI Conference on Human Factors in Computing Systems (CHI '26), April 13--17, 2026, Barcelona, Spain}
\acmPrice{}
\acmDOI{10.1145/3772318.3790748}
\acmISBN{979-8-4007-2278-3/2026/04}

\begin{document}
\title{Take the Power Back: Screen-Based Personal Moderation Against Hate Speech on Instagram}
\author{Anna Ricarda Luther}
\email{aluther@ifib.de}
\orcid{0000-0002-1169-9297}
\affiliation{%
  \institution{Institute for Information Management Bremen GmbH}
  \city{Bremen}
  \country{Germany}
}
\affiliation{%
  \institution{University of Bremen}
  \city{Bremen}
  \country{Germany}
}

\author{Hendrik Heuer}
\email{hendrik.heuer@cais-research.de}
\affiliation{%
  \institution{Center for Advanced Internet Studies (CAIS)}
  \city{Bochum}
  \country{Germany}}
\affiliation{
  \institution{University of Wuppertal}
  \city{Wuppertal}
  \country{Germany}
  }

\author{Stephanie Geise}
\email{sgeise@uni-bremen.de}
\affiliation{%
  \institution{Centre for Media, Communication and Information
Research (ZeMKI)}
  \city{Bremen}
  \country{Germany}
}
\affiliation{%
  \institution{University of Bremen}
  \city{Bremen}
  \country{Germany}
}

\author{Sebastian Haunss}
\email{sebastian.haunss@uni-bremen.de}
\affiliation{%
 \institution{Research Center on Inequality and
Social Policy (SOCIUM)}
 \city{Bremen}
 \country{Germany}}
 
\affiliation{%
 \institution{University of Bremen}
 \city{Bremen}
 \country{Germany}}

\author{Andreas Breiter}
\email{abreiter@ifib.de}
\affiliation{%
  \institution{Institute for Information Management Bremen GmbH}
  \city{Bremen}
  \country{Germany}}
  
\affiliation{%
  \institution{University of Bremen}
  \city{Bremen}
  \country{Germany}}

\renewcommand{\shortauthors}{Luther et al.}
\begin{abstract}
Hate speech remains a pressing challenge on social media, where platform moderation often fails to protect targeted users. Personal moderation tools that let users decide how content is filtered can address some of these shortcomings. However, it remains an open question on which screens (e.g., the comments, the reels tab, or the home feed) users want personal moderation and which features they value most. To address these gaps, we conducted a three-wave Delphi study with 40 activists who experienced hate speech. We combined quantitative ratings and rankings with open questions about required features. Participants prioritized personal moderation for conversational and algorithmically curated screens. They valued features allowing for reversibility and oversight across screens, while input-based, content-type specific, and highly automated features are more screen specific. We discuss the importance of personal moderation and offer user-centered design recommendations for personal moderation on Instagram.
\end{abstract}



\maketitle

\section{Introduction}

Hate speech is a well-documented issue on social media~\cite{castellanos2023hate, Castaño-Pulgarín2021Internet, Castillo-Esparcia2023Evolution}, widely affecting social media users~\cite{hateaid2021report}. In this paper, we operationalize hate speech as any form of communication that disparages individuals or groups based on characteristics such as race, gender, or religion~\cite {zhang2019hate}. Exposure to hate speech has been shown to have a silencing effect, deterring users from participating in public discourse or disclosing parts of their identities. But the consequences of hate speech online extend beyond digital spaces, posing real-world risks to mental and physical well-being~\cite{stevens2021cyber, williams2020hate}. The scope and gravity of this impact highlight the pressing need for effective strategies to mitigate hate speech and foster safer online environments.

Existing moderation systems predominantly employ platform moderation (i.e., moderation decisions apply to the entirety of the platform)~\cite{gillespie2018custodians}, which has been shown to fail to effectively combat hate speech~\cite{castellanos2023hate}. Automated hate speech detection, which is an integral part of platform moderation~\cite{gillespie2018custodians}, often lacks the context necessary to determine whether a statement is abusive~\cite{Dorn2024HarmfulSpeech} resulting in over- and undermoderation~\cite{hartmann2025lost}, particularly for content from marginalized users~\cite{Dorn2024HarmfulSpeech}. Users themselves are often aware of these contexts. Personal moderation tools can empower users to give this context, addressing this shortcoming of automated platform moderation. Personal moderation is described as tools that empower users to influence what content they see by setting their preferences e.g., through word filter or sensitivity controls, which exclusively apply to their accounts~\cite{jhaver2023users}.

In this way, personal moderation has the potential to create safer, more inclusive online spaces that respect diverse cultural and personal norms~\cite{fukuyama2020middleware, Heung2025Ignorance}. Previous work yielded insights on user preferences and design considerations for personal moderation tools~\cite{jhaver2022designing, Jhaver2023, Heung2025Ignorance}. One issue previous work identified is that personal moderation tools need to match the moderation goals of the users in granularity of control. More specifically, their findings suggest that users might have different personal moderation needs for different parts of social media apps~\cite{Jhaver2023}.

Building on this perspective, we contribute an empirical investigation of whether and how personal moderation needs vary between distinct parts of social media. To systematically investigate these differences, we introduce the theoretical construct ``screens'' (e.g., the comments section, the home feed, or the reels tab). Since personal moderation needs can be reflected in the functions users require~\cite{Jhaver2023}, we further examine personal moderation features across screens. We operationalize the term personal moderation features as discrete user interface elements that users can configure in order to avoid harmful content on their individual accounts. We address the following research questions (RQs): 
\begin{itemize}[topsep=3pt, partopsep=0pt, itemsep=3pt, parsep=0pt]
    \item Which Screens Should Provide Personal Moderation For Activists Targeted by Hate Speech?~(RQ1)
    \item What Personal Moderation Features Are Requested Per Screen by Activists Targeted by Hate Speech?~(RQ2)
\end{itemize}
To answer these research questions, we conducted a three-wave Delphi study with 40 activists living in Germany who experienced hate speech. \added{We recruited activists because their visible engagement with political issues places them at heightened risk of encountering hate speech~\cite{Meza2019Targets}, and many belong to marginalized groups that are disproportionately targeted by hate speech~\cite{Borshuk2004An}. Their extensive and often unavoidable reliance on social media to amplify their messages further increases both their exposure and the stakes of managing harassment~\cite{Momeni2017Social}.} We focused on Instagram because it is the most widely used social media platform in Germany~\cite{BeischInternetnutzung}, \added{particularly among younger and activist users~\cite{BeischInternetnutzung, Luther2025}. Prior research has also documented challenges with hate speech on Instagram, making it a relevant and high-exposure environment for studying personal moderation needs~\cite{Miranda2023Analyzing}.}

We find that personal moderation options are assessed as particularly important for screens representing conversational and algorithmically curated spaces. Although features allowing for oversight and reversibility are requested for all screens, there are also differences in the features requested for each screen. Input-based features are more frequently requested for conversational screens, where the kind of harm is known. For algorithmically curated spaces, features that are specific to different content types and features that rely heavily on automation while requiring minimal user configuration are requested.

We provide the first empirically grounded account of the screens for which activists targeted by hate speech want personal moderation~(RQ1). For each screen we provide the features that are considered most important~(RQ2). In the discussion, we translate the identified user needs into concrete design recommendations for personal moderation tools. Together, these contributions advance research on personal moderation on social media and provide actionable pathways for designing moderation practices that are both scalable and responsive to the needs of those affected by hate speech.

\section{Background} 

\subsection{Different Types of Moderation on Social Media}
We situate this work within existing scholarship that differentiates between different moderation approaches. \citeauthor{Jhaver2023} (\citeyear{Jhaver2023}) categorized moderation approaches on social media based on the scope of the respective moderation decision, resulting in the following three categories: platform moderation (moderation decisions that apply to the entirety of the platform), community moderation (moderation decisions that apply to specific communities), and personal moderation (moderation decisions that apply to the view of an individual user). They further distinguish between moderation decisions that are based on the content (e.g., removing content because it contains a specific keyword) and moderation decisions that are based on the account (e.g., banning a user because they repeatedly violated platform guidelines)~\cite{Jhaver2023}. Given that community moderation possibilities on Instagram are limited to individual channels of content creators~\cite{instagram_privacy_settings}, in the following, we focus on how personal moderation can address limitations of platform moderation.

\subsection{Limitations of Platform Moderation}
\label{PlatformModeration}

In the following, we lay out the limitations of platform moderation and highlight how personal moderation can, in part, address them. One key limitation of platform moderation is its one-size-fits-all approach. While there is undeniably content on social media that merits universal removal due to its explicit harm (e.g. child sexual abuse material~\cite{gillespie2018custodians}), for a large portion of content, this distinction is not that clear-cut~\cite{magdy2025should}. What constitutes hate speech depends on norms, identities, and contexts that differ between cultures, communities, and individuals~\cite{jiang2021understanding, weld2022makes, Seering2019Moderator, jhaver2018online}. With a user base spanning billions of users across the major social media platforms~\cite{statista2024socialmedia}, platform-wide moderation is constrained in its ability to accommodate the resulting diverse expectations of what constitutes acceptable speech. Personal moderation could add more nuance to this one-size-fits-all approach by integrating the individual user perspective into the moderation process.

Further, social media platforms predominantly employ automated systems to enable content-based platform moderation~\cite{gillespie2018custodians}. Previous research has repeatedly shown that automated hate speech detection systems show bias against marginalized communities~\cite{Poletto2020Resources}. This demonstrates the layered nature of marginalization: those most often targeted by hate speech are simultaneously disadvantaged by biased systems, compounding the harms they face online. The biases in automated hate speech detection are evident in two ways: through undermoderating hate speech directed at marginalized groups~\cite{Arshad-Ayaz2022Perspectives, Retta2023A} while simultaneously overmoderating counter speech and reappropriation of slurs~\cite{Dorn2024HarmfulSpeech, dixon2018measuring}. Reappropriation refers to the practice of marginalized groups to use derogatory terms referring to that group as self-description, constituting a form of empowerment or resistance~\cite{Cervone2020The}.

Improving the performance of automated hate speech detection remains difficult due to several key challenges. For example, it remains an open question how to meaningfully incorporate the context necessary to accurately detect hate speech ~\cite{Dorn2024HarmfulSpeech}. Users themselves are, however, often aware whether user-generated content constitutes counter speech or whether a term is reappropriated. Personal moderation would allow the users to give this context and could thus effectively address this shortcoming of platform moderation. 

Automated hate speech detection tools used for platform moderation further struggle with data drift, where the underlying data distribution changes over time~\cite{Dong2024Efficiently}, concept drift, where the relationship between input data and the target labels shifts~\cite{Bayram2022From}, and model drift, where the model's performance degrades due to evolving patterns or societal norms~\cite{Manias2023Model}. These dynamics are critical, since new slurs or code words emerge and the meaning of existing terms can change over time~\cite{tahmasbi2021go}. Further, offenders can intentionally obfuscate or alter spelling to evade detection~\cite{grondahl2018all}. Thus, pre-trained models are bound to lag behind these developments. The use of sarcasm, irony, or coded language further complicates the issue, leading to worse detection accuracy~\cite{hartmann2025lost, perez2023assessing}. As the repeated detection of harmful content serves as a prerequisite for account-based platform moderation in the majority of platform guidelines~\cite{meta_restricting_accounts, tiktok_content_violations_bans, reddit_content_policy}, it suffers from the same limitations. This motivates us to focus on personal moderation. By empowering users to control moderation, the efficacy of moderation efforts can more easily be adapted to evolving language and norms.

In sum, platform moderation struggles with biases and a lack of contextual awareness, disproportionately harming the very communities most frequently targeted by hate speech. These limitations highlight the need for complementary approaches such as personal moderation, which can integrate individual user needs into moderation processes.

\subsection{Personal Moderation}
Both scholars and platform designers have begun to recognize personal moderation as an important direction for improving moderation on social media. In a nationally representative survey of 984 US adults, \citeauthor{jhaver2023users} (\citeyear{jhaver2023users}) have shown that users prefer personal moderation tools over platform-directed moderation for norm-violating content on social media. Scholars have also increasingly called for personal moderation solutions~\cite{Jhaver2023, fukuyama2020middleware, noauthor_middleware_nodate} to improve moderation practices on social media. As a response, social media platforms have incorporated personal moderation features and several third-party actors have developed personal moderation tools. Examples include Gobo Social~\cite{mit_gobo}, Bodyguard~\cite{bodyguard_ai}, Block Party~\cite{blockparty_app}, and FilterBuddy~\cite{jhaver2022designing}. FilterBuddy, for example, addresses weaknesses in rule-based word filters, which require users to manually enter every linguistic variant of a derogatory term and to anticipate an often overwhelming range of terms~\cite{jhaver2022designing}. FilterBuddy reduces this burden by automatically generating variants, enabling the import of group-specific keyword lists, and allowing list sharing between users~\cite{jhaver2022designing}. 

\subsubsection{Screens as a Lens for Granularity of Control in Personal Moderation}
Despite the high importance of personal moderation both for users~\cite{jhaver2023users, Heung2025Ignorance} and scholars~\cite{Jhaver2023, fukuyama2020middleware, noauthor_middleware_nodate} and the increasing implementation of personal moderation functionalities, open questions remain regarding to the design of personal moderation tools. One critical gap is the necessary granularity of control of such tools. Previous work showed that participants experienced harm across different parts of social media (such as news feeds, direct messages, or user profiles), and suggested that user needs for personal moderation might differ between these parts~\cite{Jhaver2023}. For instance, some users may want to always filter out particular words in their news feed but not in their direct messages. These insights motivated us to empirically examine whether user needs vary across different parts of the platform, which we formalize in terms of screens: For which screens do activists targeted by hate speech want personal moderation? (RQ1)

\subsubsection{Key Features for Personal Moderation}
Based on the assumption that personal moderation needs might differ between screens, this raises the questions what features are requested for the respective screens. Prior research has begun to explore what features users want from both account-based and content-based personal moderation~\cite{Heung2025Ignorance, Jhaver2023}. For example, interviews evaluating existing moderation tools and a technology probe exploring new personal moderation options revealed key insights into the kind of features users need~\cite{Jhaver2023}. Similarly, a design probe for disabled social media users highlighted the need for configurable filtering of ableist content~\cite{Heung2025Ignorance}. While this prior work offers a strong foundation, it has not linked these feature requests to screens nor has it enabled users to rate and hierarchize these features to identify clear priorities. Prioritizing these features is crucial to focus the design of personal moderation tools on the most impactful and widely desired solutions. Employing the Delphi method allowed us to do this~\cite{hsu2007delphi, cuhls2023delphi} and gain insights into our second research question: What personal moderation features do activists targeted by hate speech request across screens? (RQ2) 

\section{Methods}
\subsection{The Delphi Method}
The Delphi method is a structured, multi-round, path-dependent survey approach designed to build consensus among a group of experts~\cite{Wilson2024Delphi}. We chose the Delphi method for our study because it allows participants to iteratively refine their judgments across multiple rounds. The Delphi method is path-dependent in the sense that items are added based on prior anonymized and aggregated participant input~\cite{cuhls2023delphi}. Items that do not receive a previously specified threshold of agreement can also be removed~\cite{hsu2007delphi}. Its anonymous format limits the influence of social dynamics on the outcome~\cite{cuhls_argumentative_2024, hsu2007delphi}. Furthermore, Delphi studies are often employed in technology forecasts to address emerging design spaces, where consensus-building can meaningfully inform future development~\cite{alarabiat_delphi_2019, Wilson2024Delphi}. Finally, by its funnel-like structure, the Delphi process yields stable priorities that can inform platform design~\cite{cuhls2023delphi}. The structured, iterative, and consensus-oriented approach allows us to contribute a systematic assessment of which personal moderation features users request based on the screens on which they are applied. 

\subsection{Study Procedure}

We conducted a three-wave Delphi study with 40 activist social media users from Germany targeted by hate speech. Each wave was conducted as an online survey administered via SoSci Survey. Before launching the study, we conducted three iterations of informal pre-tests with eight colleagues and graduate students from our institution. They provided feedback on question formulation, clarity of instructions, and survey flow, which we incorporated into each iteration, leading to the final design. 

Local regulations do not require formal institutional ethics review for this type of research. Nonetheless, the institutional review board (IRB) of the second author's institution reviewed and approved the study proposal.  

Each wave served a specific purpose: Wave~1 offered an initial assessment of the personal moderation tool and extended the response possibilities based on free user input, Wave~2 provided a validation of this initial assessment, and Wave~3 provided a ranking of the most important screens and features. An overview of the study procedure can be found in Table \ref{tab:StudyProcedure}.

\begin{table*}[t!]
\Description{Table 1 summarizes the procedures used in each of the three Delphi study waves. The table has three columns (Wave 1, Wave 2, and Wave 3), each containing a short list describing the steps participants completed in that phase.
Wave 1 describes the baseline procedure: participants read a scenario, rated the importance of all screens, and for screens they considered important, rated three predefined features and added new feature requests through free-text input.
Wave 2 follows the same structure but with two changes: all questions about the Search Results screen were removed because it did not receive majority support in Wave 1, and participants rated a larger set of new feature requests generated from Wave 1. Median ratings from Wave 1 were displayed alongside existing items.
Wave 3 again built on the prior procedure but shifted from rating features to ranking them. Additional feature requests from Wave 2 were included, and median ratings from Wave 2 were shown next to items already present. Overall, the table highlights how each wave added or refined tasks to further narrow and prioritize user-requested features.}
\caption{Overview of the study procedure across the three Delphi waves. All screens that were presented in Wave 1 are explained in Table \ref{IntroductionScreens}}
\label{tab:StudyProcedure}
\centering
\begin{tabular}{p{0.33\textwidth} p{0.33\textwidth} p{0.33\textwidth}}
\toprule
\textbf{Wave 1 (W1)} & \textbf{Wave 2 (W2)} & \textbf{Wave 3 (W3)} \\
\midrule
The participants completed the initial survey:
\begin{enumerate}[leftmargin=*]
    \item The participants read the scenario.
    \item The participants \textbf{rated the importance} of all \textbf{screens} and explained their ratings.
    \item For all screens rated as important: 
        \begin{enumerate}
            \item Participants \textbf{rated the importance of 3 features} (from literature).
            \item Participants \textbf{added feature requests} via free text input.
        \end{enumerate}
\end{enumerate}
&
The participants followed the same procedure as in W1, but:
    \begin{enumerate}
        \item All questions relating to the screen \textbf{\textit{Search Results} were removed}. (The screen did not receive the majority vote in W1).
        \item Participants also \textbf{rated further feature requests} (37 features were added based on free text input from W1).
        \item All \textbf{median ratings from W1} were shown next to items already present in W1.
    \end{enumerate}
&
The participants followed the same procedure as in W2, but:
    \begin{enumerate}
        \item Participants did not rate the features, they \textbf{ranked} the features. 
        \item Participants also \textbf{ranked further feature requests} (6 features were added based on free text input from W2).
        \item All \textbf{median ratings from W2} were shown next to items already present in W2.
    \end{enumerate}

\\
\bottomrule
\end{tabular}
\end{table*}

In Wave~1, participants received a scenario:

\begin{quote}
    ``\textit{Imagine a social media app like Instagram. Hate speech is often not removed from this app. As an experienced user, you know the challenges this poses. Now you have the opportunity to design a new tool that gives you more control over how hate speech is filtered on your account. Imagine you could decide how this tool would work. We would like to know how you envision this tool. Think about your experiences with hate speech and consider how this tool could protect you from hate speech on social media. You don't need to think about whether your ideas and wishes are technically feasible.}'' (translated from German). 
\end{quote}

Based on this scenario, we asked the participants to rate the importance of different screens across the interface of Instagram on a 5-point Likert Scale~(1 = unimportant, 2 = somewhat unimportant, 3 = neither, 4 = somewhat important, 5 = important, -1 = don't know), to understand which screens users consider most important for controlling content filtering~(RQ1). Participants were asked to explain their importance ratings per screen. 

Each screen was explained with a short description and a screenshot to ensure that participants understood each screen in the same way. All screens, their explanations, and their labels are listed in Table~\ref{IntroductionScreens}. To support comprehension, we include images of the respective screens in the results in Figure \ref{fig:overview_screens}.

\begin{table*}[t!]
    \Description{Table 2 provides an overview of the different screens included in the analysis. The table has two columns and ten rows. The first column, Screen Label, lists the names of the screens, while the second column, Explanation, describes the type of content each screen includes.
    The listed screens and their descriptions are: Home Feed, when the user scrolls the Home Feed; Explore Tab, all content that is visible in the Explore Tab; All Comments, all comments on all kinds of content; Comments on Specific Posts, all comments on individual posts; Comments on Own Posts, all comments that address their own content; All Direct Messages, all direct messages towards the respective user; Direct Messages from Specific Accounts, all direct messages towards the respective user from individual users; Search Results, all content that is shown based on search entries; Reels Tab, all content visible on the Reels Tab; and Optional Custom Level, possibility to enter further screens.}
    \caption{This table provides an overview of all screens, including their descriptions and labels used in the analysis. They are listed in the order in which they are presented to the participants. Participants rated the importance of personal moderation on each screen on a 5-point Likert Scale in all three waves.}
    \begin{tabular}{l l}
    \toprule
    \textbf{Screen Label} & \textbf{Explanation} \\
    \midrule
    \textit{Home Feed} & When the user scrolls the Home Feed \\     
    \textit{Explore Tab} & All content that is visible in the Explore Tab\\
    \textit{All Comments} & All comments on all kinds of content\\
    \textit{Comments Specific Posts} & All comments on individual posts\\ 
    \textit{Comments Own Posts} & All comments that address their own content\\
    \textit{All Direct Messages} & All direct messages towards the respective user\\
    \textit{Direct Messages from Specific Accounts} & All direct messages towards the respective user from individual users\\
    \textit{Search Results} & All content that is shown based on search entries\\
    \textit{Reels Tab} & All content visible on the Reels Tab\\
    \bottomrule
    \end{tabular}
    \label{IntroductionScreens}
\end{table*}

Only when a participant rated a screen as neutral, rather important, or important, we asked the participant to rate features of the hypothetical personal moderation tool for this screen. We then asked the participant to suggest additional features for this screen. Three features derived from the literature are presented to the participants to help them concretize their ideas. These examples were not meant to constrain responses but to stimulate reflection and ensure that all participants had a shared starting point for discussion, as is common in Delphi studies~\cite{cuhls_argumentative_2024}. The first option is \textbf{Filtering Based on Words}. This feature would allow the user to enter specific strings. Content that explicitly contains these strings is filtered~(e.g., slurs or keywords). This feature was chosen as initial input, because it is both something already implemented on some social media platforms~\cite{instagram_privacy_settings, facebook_hate_speech_help}, while also being criticized for its limitations~\cite{jhaver2022designing}. The second option is \textbf{Filtering Based on Topics}. This feature would enable the user to enter specific topics, and then the content related to these topics would be filtered. In distinction to the \textbf{Filtering Based on Words}, this does not just mean string matching, but also includes synonyms, semantically similar words, or that the content deals with the broader topic the user entered. This feature was chosen as initial input, because it has been explored in prior studies for their potential to reduce exposure to harmful content~\cite{salminen_taking_2021}. However, the differences in importance of a topic filter between screens remain an open question. The third feature is \textbf{Customizing the Response to Detected Hate Speech}. Users can decide how detected hate speech is treated: whether it is removed, blurred, hidden behind a warning, or moved to a ``review later'' folder. The choice to include this feature as initial input is based on previous work highlighting that users' fear of missing out both negatively impacts their evaluation of personal moderation~\cite{Jhaver2023, Heung2025Ignorance} and even negatively influences the adoption of account-based personal moderation tools~\cite{jhaver2025personal}. Participants were then encouraged to suggest additional features for each screen through free text input. 

As a core element of a Delphi study, the procedure for Wave~2 was the same as for Wave~1, but with three key modifications. First, we focused on the screens and features supported by the majority ($\geq$ 50\% of participants rated the screen or feature with a Likert rating $\geq$ 4). Second, we displayed the median importance rating of Wave~1 for each screen to the participants to provide feedback on the group’s collective perspective, fostering reflection~\cite{cuhls2023delphi}. Third, we expanded the list of personal moderation features by integrating suggestions from participants' free-text responses. This design encouraged participants to reflect, reassess their initial input, and engage with new filtering ideas emerging from peer feedback. 

To produce a ranking of personal moderation features, Wave~3 focused on ranking the personal moderation features. We only included screens and features that were supported by a majority in Wave~2~($\geq$ 50\% of participants with a Likert rating $\geq$ 4). We compiled the ranking based on a point allocation task~\cite{collewet_preference_2023}. For example, for the screen \textit{All Comments}, we presented 14 features and participants could distribute 14 points. Thus, they could either give all points to one feature, distribute the points equally among all features, or anything in between. This approach forced participants to weigh trade-offs between features, ensuring that the final ranking reflects relative priorities rather than isolated importance judgments. We turned the points into percentages to make them comparable and ranked the features based on the percentage.

\subsection{Recruitment \& Sample}

We recruited activists who regularly use social media and have encountered hate speech online. We operationalize the term ``activist'' as individuals who actively participate in social movement organizations. To be included in the study, participants had to (1) be involved in a social movement, (2) have used social media at least twice in the past month specifically for activities related to their movement, and (3) have experienced hate speech online.

We followed a multi-step recruitment strategy. We began by emailing social movements and organizations, based on a systematic collection of all social movements in three medium-to-large German cities characterized by active civic participation, diverse political landscapes, and established grassroots networks. We also recruited participants through contacts from previous collaborations with social movements. Further, we employed snowball sampling. 

We observed the following participant attrition: 58 participants completed Wave 1, 47 completed Wave 2, and 40 completed Wave 3. \added{While free-text inputs from all participants in each wave were used to inform and expand the set of feature suggestions in subsequent waves, the analysis of ratings and rankings was restricted to the 40 participants who completed all three waves. This decision was made to maintain a consistent panel across waves, ensuring that observed differences in ratings reflect within-participant changes over time rather than changes in panel composition. The following sample description, therefore, refers only to participants who completed all three waves.} 
The mean age of the participants is 32.3 years (SD~=~9.5), the youngest participant was 20 years old, and the oldest was 63 years old. The sample consists of 16 women, 12 men, and 11 non-binary individuals. One participant did not disclose their gender. Notably, the proportion of non-binary participants is comparatively large in relation to similar studies~\cite{Jhaver2023}, \added{likely because we recruited within social movements concerned with LGBTQIA+ activism}. This representation offers valuable insights from a group that is often underrepresented in empirical work~\cite{Spiel2019Nonbinary}. The panel reflects diverse activist backgrounds, allowing us to capture perspectives from activist populations, as done by previous work~\cite{Elmimouni2025Exploring, Luther2025}. Participants could select multiple terms describing their activism or give free text inputs describing their political activism. The sample included individuals engaged in left political activism~(14), LGBTQIA+ activism~(11), climate and environmental justice~(8), feminism and women’s rights~(5), ethnic and religious minority advocacy~(5), anti-fascism~(4), refugee advocacy~(2), anti-ableism~(1) and digital safety~(1).

Each participant received €20~(\$22.60) for completing each Wave of the study, which took them less than 30 minutes on average (SD~=~5.3).

\subsection{Analysis}
We inspected all responses from participants who completed all three waves and checked for completeness and straightlining patterns. We also examined whether participants provided free-text input that indicated they understood what was being asked, e.g. by providing explanations or examples relevant to the question. All responses met these criteria, and therefore, no cases required removal.

We applied reflexive thematic analysis following Braun \& Clarke to the free-text responses~\cite{braun2019reflecting}. This is done to develop new feature suggestions for Wave~2 and Wave~3~(RQ2). The first author engaged in repeated readings of the data, iteratively navigating the content to gain familiarity. The first author followed an open coding process, developing initial codes from the data. This inductive coding was followed by axial coding~\cite{corbin2014basics}. Code development was discussed collaboratively with the second author in six revision sessions until all disagreements were resolved. Through an iterative process of clustering, refining, and combining codes, we identified categories that represent novel filtering options~(RQ2). For example, we split the code \textit{Multimedia Filter} into subcodes such as \textit{Filtering Based on Image Content} or \textit{Filtering Based on Audio Content} while we merged the codes \textit{Filtering Aggressive Tone} and \textit{Filtering Condescending Tone} into \textit{Filtering Based on Tone}. 

We present the quantitative data as the percentage of participants who rated the screen~(RQ1) or the feature~(RQ2) as important (Likert ratings $\geq$ 4). For a hierarchization of the concrete feature suggestions from Wave~3, we analyzed the ranks of the feature requests, which are based on the percentage of points each feature received. In the paper, we focus on all features that have more than 5\% of all points.

\section{Results}

\subsection{RQ1: Which Screens Should Provide Personal Moderation For Activists Targeted by Hate Speech?}
\label{RQ1}

To gain insights into the screens for which activist users targeted by hate speech want personal moderation, we asked participants to rate the importance of personal moderation across various screens. Seven out of ten participants rated all screens as important (Likert ratings $\geq$ 4) except for the screen \textit{Search Results}. This screen was the only screen not rated as important by the majority of participants and was subsequently removed from Wave~2 and Wave~3. A table with all ratings of all screens across all waves can be found in the Appendix (Table ~\ref{tab:AllInteractionContexts}). In the following, the results of Wave~3 are presented.

\begin{figure*}[t!]
    \centering
    \includegraphics[width = \textwidth]{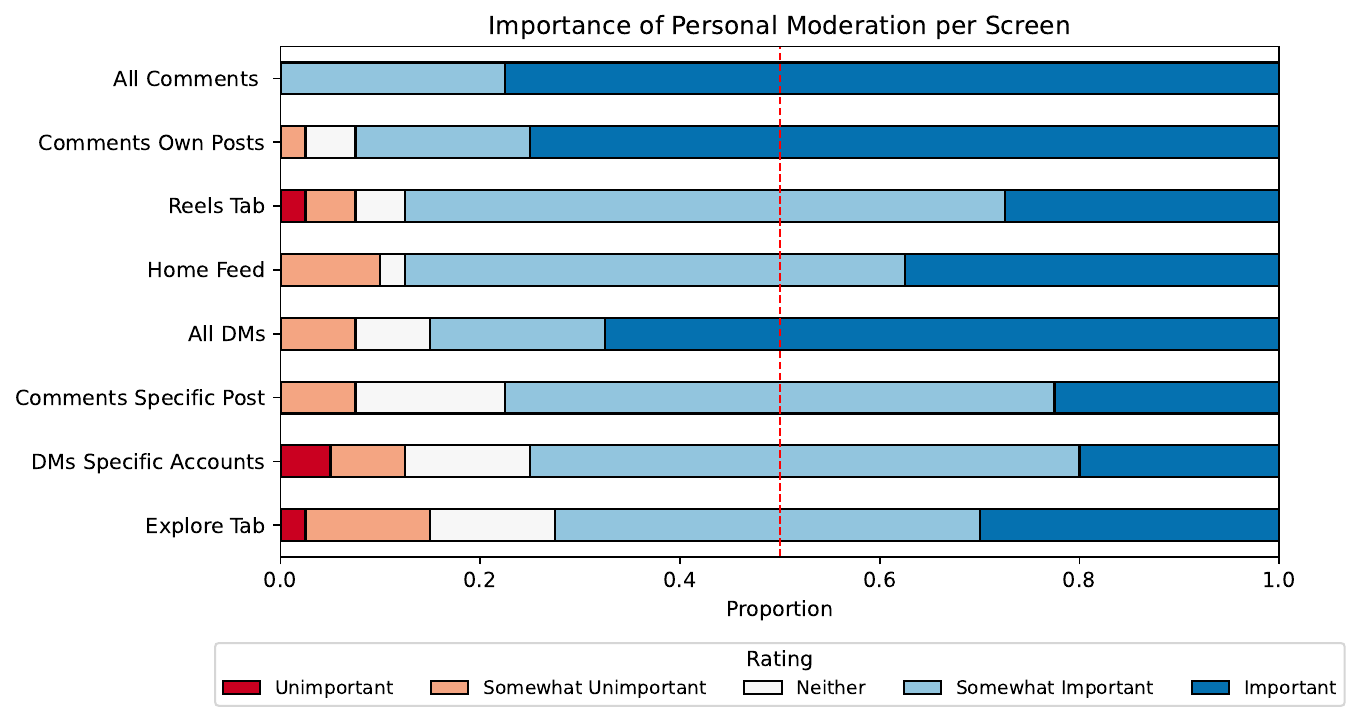}
    \Description{Figure 1 is a horizontal stacked bar chart showing the distribution of importance ratings for personal moderation across eight different social media screens among 40 activist social media users who experienced hate speech. The screens, shown along the y-axis, are: All Comments, Comments on Own Posts, Reels Tab, Home Feed, All DMs, Comments on a Specific Post, DMs with Specific Accounts, and Explore Tab. Each bar represents one screen, with colored segments indicating the proportion of participants who selected each of the five ratings. The ratings are Unimportant, Somewhat Unimportant, Neither, Somewhat Important, and Important. The x-axis shows proportions from 0 to 1, and a vertical dashed line marks the 50\% point. Overall, “All Comments” and “Comments on Own Posts” have the highest proportions of participants rating moderation as important, while the “Explore Tab” and “DMs Specific Accounts” have the lowest. Most screens show a majority of responses in the “Somewhat Important” or “Important” categories. Only “Reels Tab”, “DMs Specific Accounts”, and “Explore Tab” were labeled as “Unimportant” by some participants.}
    \caption{The distribution of importance ratings across all screens in Wave~3 among 40 activist social media users who experienced hate speech directed at them. The vertical red dashed line indicates the 50\% mark, showing which ratings are shared by the majority of participants. The screens are sorted by the share of participants who rate it as important.}
    \label{fig:InteractionContextsW3}
\end{figure*}

A strong majority of at least eight out of ten participants rated \textit{All Comments} (100\%), \textit{Comments Own Posts} (92.5\%), \textit{Home Feed} (87.5\%), \textit{Reels Tab} (87.5\%) and \textit{All DMs} (85.0\%) as important. Among those five screens, three screens describing conversational spaces were highlighted by many participants as especially important for personal moderation: At least six in ten participants rated \textit{All Comments} (77.5\%), \textit{Comments Own Posts} (75.0\%), and \textit{All DMs} (67.5\%) as highly important. Less than four in ten participants rated the two screens describing algorithmically curated spaces, namely the \textit{Reels Tab} (27.5\%) and the \textit{Home Feed} (37.5\%) as highly important.

For the screens \textit{All Comments, Comments Own Post} and \textit{All DMs}, participants described the high exposure to hate speech in the comments as a key motivation for personal moderation on these screens (e.g., P2, P17, P21, P25, and P39). 
At the same time, each screen revealed distinct motivations: for \textit{All Comments}, P16 highlighted the lack of control over ``who I see in the comments''. P16 directly contrasts this with the \textit{Home Feed}, where they use the ``Following'' Tab to ensure that only content from followed accounts is shown. This kind of control is lacking for the comments. For \textit{Comments Own Post} P31 noted that this screen ``\textit{directly affects [their] mental space and sense of safety}'', a concern shared by several participants (e.g., P25, P32, and P36).

Participants also highlighted distinct motivations for personal moderation on the algorithmically curated screens \textit{Reels Tab} and the \textit{Home Feed}. The participants emphasized that the \textit{Home Feed} is the entry point for all interactions on a social media app and is, therefore, a particularly important screen for personal moderation (e.g., P31, P32, and P39). For the \textit{Reels Tab}, participants emphasized the importance of personal moderation because the Reels ``\textit{are autoplayed without a preview of what the reel is about}'' (P27). Also, the large amount of time spent on the \textit{Reels Tab} is reported as a reason for the high importance rating for this screen (e.g., P27, P32, and P38). The limited control over the kind of content displayed on the \textit{Reels Tab} is another reason for the importance of personal moderation applying to the \textit{Reels Tab} (P39).

Three screens are rated as important by less than 80\% of the participants: \textit{Comments Specific Post} (77.5\%), \textit{DMs Specific Accounts} (75.0\%), and \textit{Explore Tab} (72.5\%). P11 explained that they rated the importance of personal moderation for \textit{Comments Specific Posts} and \textit{DMs Specific Accounts} as lower because they assumed it to be ``\textit{quite complicated for the person filtering.}'' Others described that this degree of control was too fine-grained, and they preferred the same personal moderation for all comments and all DMs (e.g., P7, P11, P17, and P37). For the screen \textit{DMs Specific Accounts}, P31 reported that ``\textit{the risk of [being targeted by] hate speech or harassment is higher from specific individuals or types of accounts (e.g., strangers, repeat offenders, public reply trolls)}'', others highlighted how they ``\textit{wouldn't be able to select specific accounts in advance}'' (P7 and P15). For the \textit{Explore Tab}, participants (P1, P3, and P25) explained that they rarely use this screen, especially ``\textit{for political accounts}'' (P25), which reduced the importance of personal moderation on that screen for them. 

\subsection{RQ2: What Personal Moderation Features Are Requested Per Screen by Activists Targeted by Hate Speech?}
\label{RQ2}

\begin{figure*}[t]
\centering
\begin{subfigure}[t]{0.225\textwidth}
    \centering
    \includegraphics[width=\linewidth]{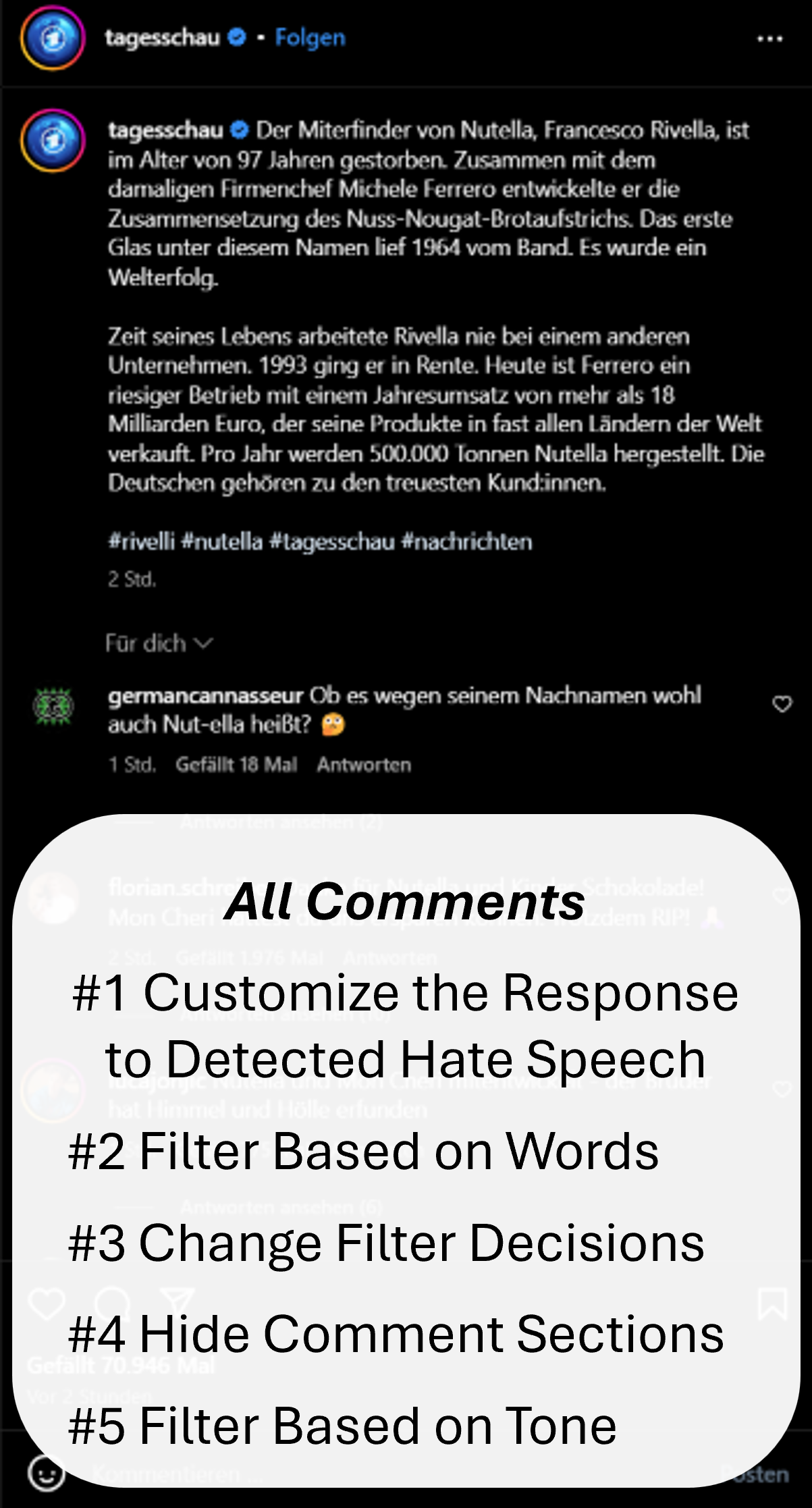}
    \Description{Figure 2a is a screenshot of the comment section of an Instagram post from the verified account “tagesschau.” The screenshot shows the post description at the top and user comments below it. Overlaid in the lower area of the image is a white, rounded box labeled Comments Own Posts. The five highest-ranked personal moderation features listed inside the box are: (1) Customize the Response to Detected Hate Speech, (2) Filter Based on Words, (3) Change Filter Decisions, (4) Hide Comments Sections, and (5) Filter Based on Tone.}
    \caption{\textit{All Comments}}
    \label{fig:AllComments}
\end{subfigure}
\hfill
\begin{subfigure}[t]{0.225\textwidth}
    \centering
    \includegraphics[width=\linewidth]{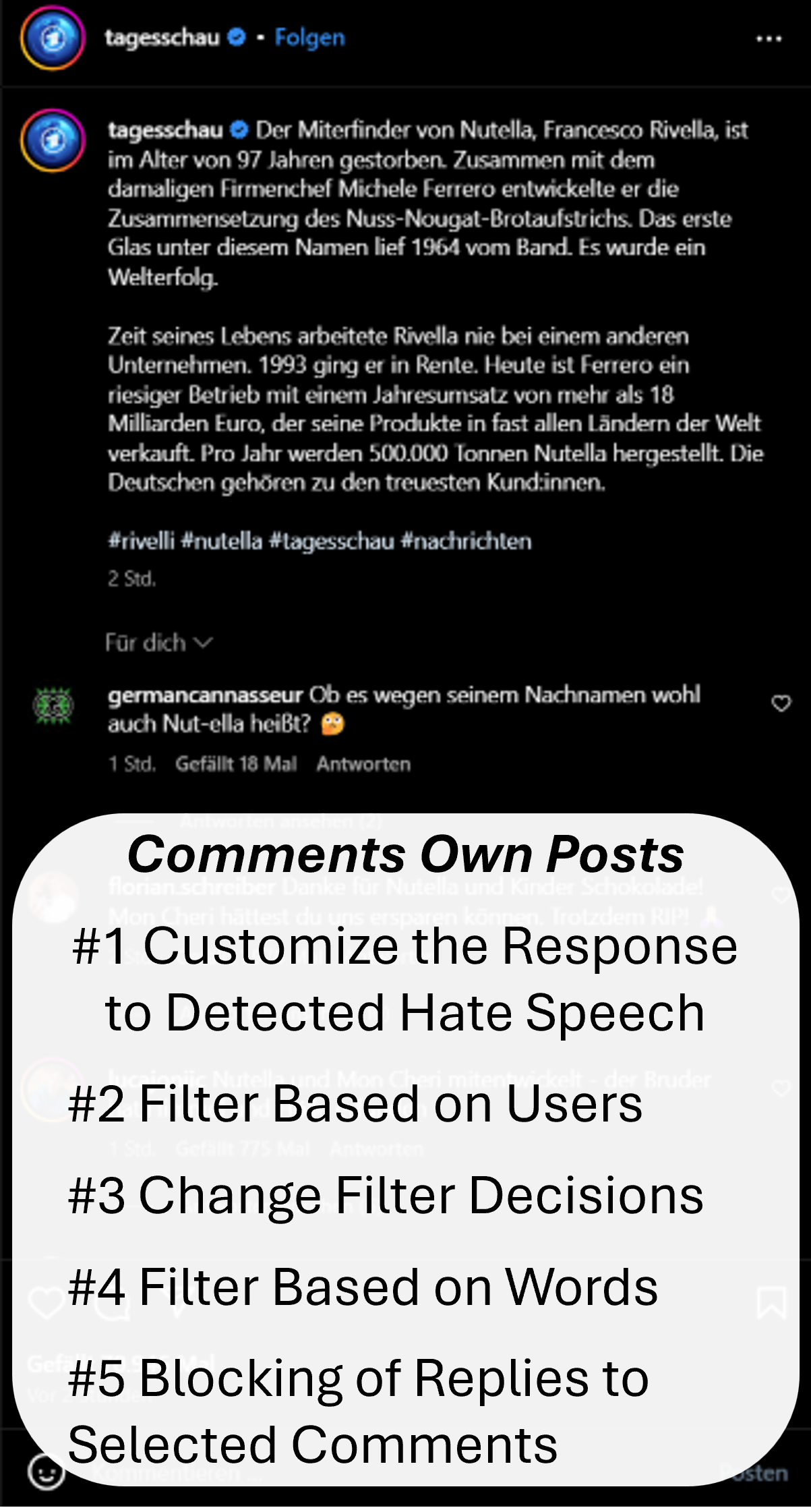}
    \Description{Figure 2b is a screenshot of the comment section of an Instagram post from the verified account “tagesschau.” The screenshot shows the post description at the top and user comments below it. Overlaid in the lower area of the image is a white, rounded box labeled Comments Own Posts. The five highest-ranked personal moderation features listed inside the box are: (1) Customize the Response to Detected Hate Speech, (2) Filter Based on Users, (3) Change Filter Decisions, (4) Filter Based on Words, and (5) Blocking of Replies to Selected Comments.}
    \caption{\textit{Comments on Own Posts}}
    \label{fig:owncomments}
\end{subfigure}%
\hfill
\begin{subfigure}[t]{0.225\textwidth}
    \centering
    \includegraphics[width=\linewidth]{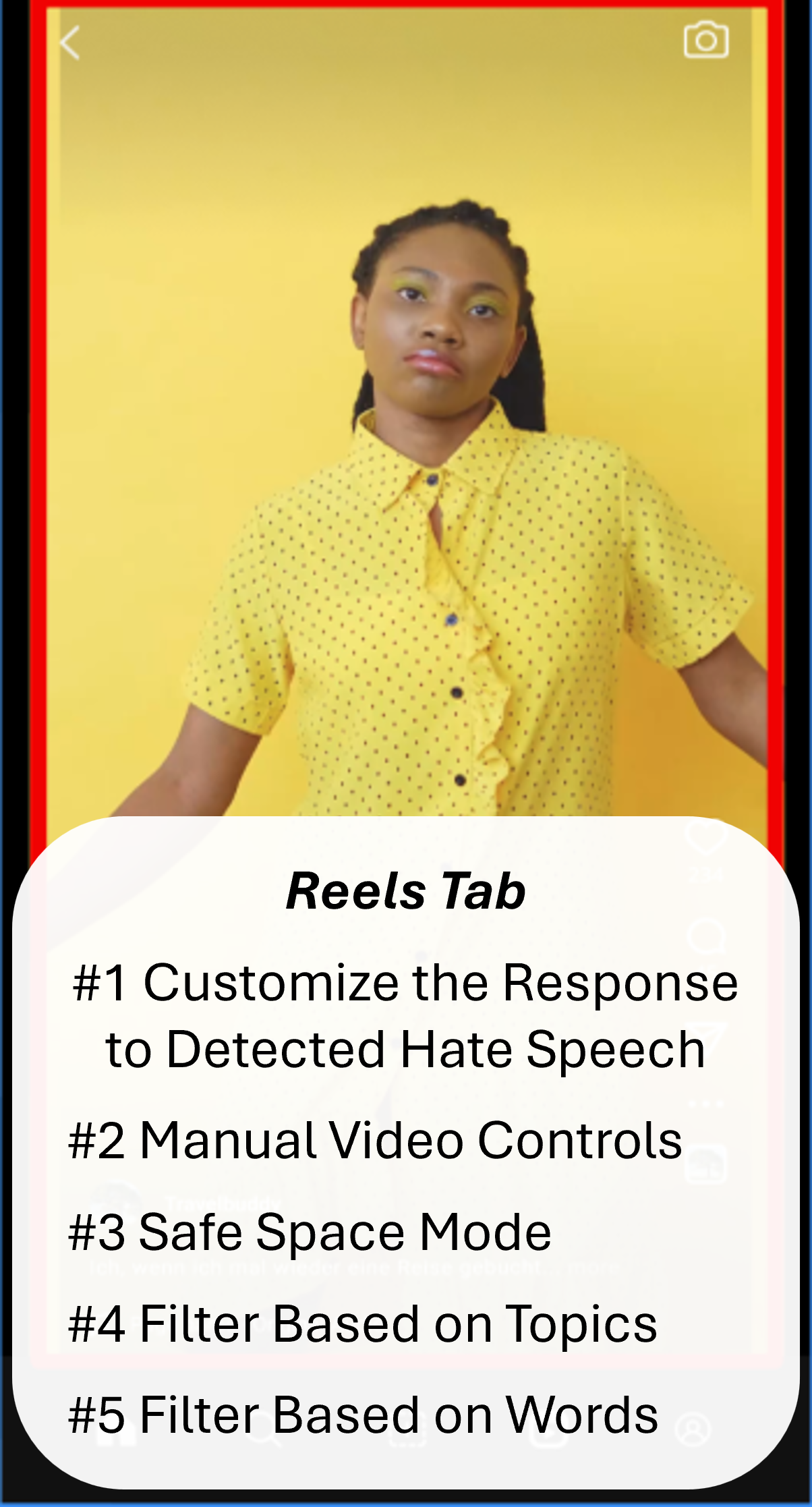}
    \Description{Figure 2c is a screenshot of an Instagram Reel showing a person standing against a yellow wall. Overlaid in the lower area of the image is a white, rounded box labeled Reels Tab. The five highest-ranked personal moderation features listed inside the box are: (1) Customize the Response to Detected Hate Speech, (2) Manual Video Controls, (3) Safe Space Mode, (4) Filter Based on Topics, and (5) Filters Based on Words.}
    \caption{\textit{Reels Tab}}
    \label{fig:reels}
\end{subfigure}

\vspace{0.6em} 

\begin{subfigure}[t]{0.225\textwidth}
    \centering
    \includegraphics[width=\linewidth]{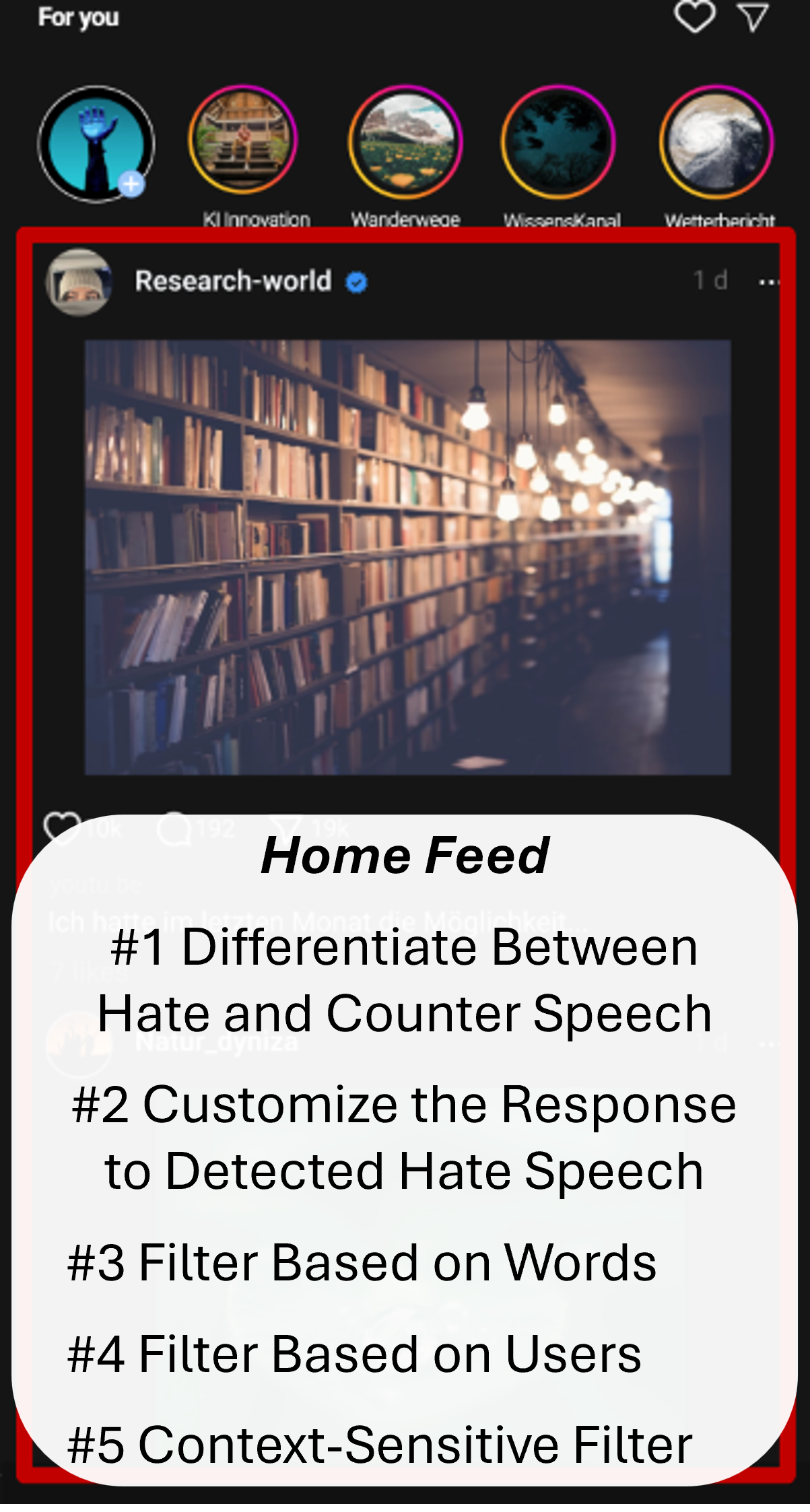}
    \Description{Figure 2d is a screenshot of an Instagram Home Feed showing a post with an image of a library filled with books and hanging lights. Overlaid in the lower area of the image is a white, rounded box labeled Home Feed. The five highest-ranked personal moderation features listed inside the box are: (1) Differentiate Between Hate and Counter Speech, (2) Customize the Response to Detected Hate Speech, (3) Filter Based on Words, (4) Filter Based on Users, and (5) Context-Sensitive Filter.}
    \caption{\textit{Home Feed}}
    \label{fig:homefeed}
\end{subfigure}%
\hspace{0.125\textwidth}
\begin{subfigure}[t]{0.225\textwidth}
    \centering
    \includegraphics[width=\linewidth]{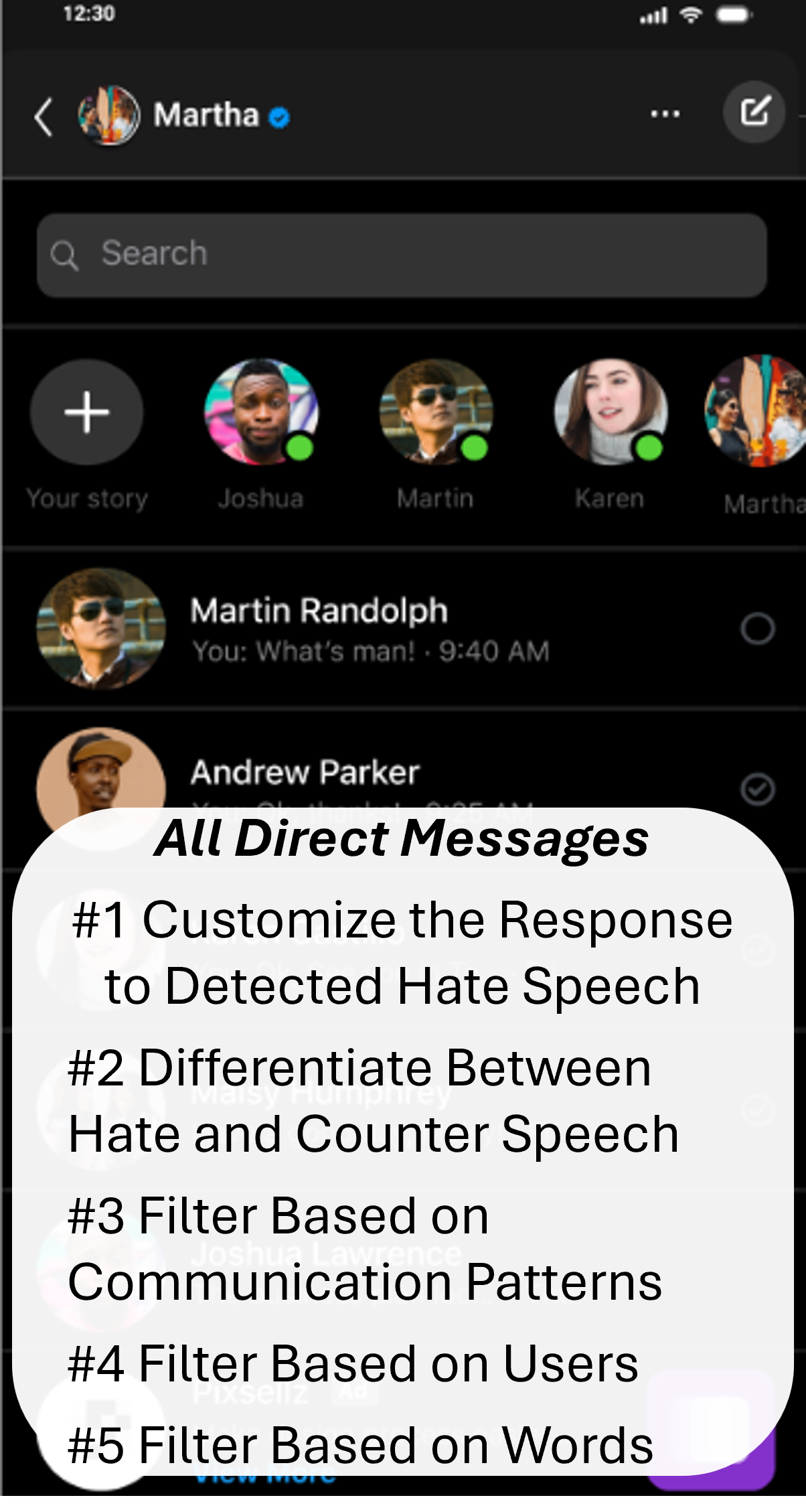}
    \Description{Figure 2e is a screenshot of the DMs section of Instagram, showing the top navigation bar with a username and a list of recent conversations below. Overlaid in the lower area of the image is a white, rounded box labeled All Direct Messages. The five highest-ranked personal moderation features listed inside the box are: (1) Customize the Response to Detected Hate Speech, (2) Differentiate Between Hate and Counter Speech, (3) Filter Based on Communication Patterns, (4) Filter Based on Users, and (5) Filter Based on Words.}
    \caption{\textit{All DMs}}
    \label{fig:alldms}
\end{subfigure}

\caption{This figure shows the example images shown to the participants for the screens \textit{All Comments} (\ref{fig:AllComments}), \textit{Comments Own Posts} (\ref{fig:owncomments}), \textit{Reels Tab} (\ref{fig:reels}), \textit{Home Feed} (\ref{fig:homefeed}) and \textit{All DMs} (\ref{fig:alldms}). On top, the five highest-ranked personal moderation features for the respective screens are displayed, arranged according to the participant ranking.}
\label{fig:overview_screens}
\end{figure*}

We examine feature requests in relation to different screens. Participants indicated their feature requests for personal moderation separately for each screen. We report the importance ratings of the features for Wave 1 (W1) and Wave 2 (W2), and the final ranking in Wave 3 (W3). The screens are presented in order of their perceived importance (see Section \ref{RQ1} for reference and Figure \ref{fig:overview_screens} for a visual representation of these five screens). The feature requests, the ratings and rankings for the screens not discussed in the following can be found in the Appendix (Tables \ref{tab:CommentsSpecificPosts} - \ref{tab:ExploreTab}). 

\subsubsection{Screen: All Comments}

\begin{table*}
    \centering
    \Description{Table 3 presents the final Wave 3 ranking of feature requests for the “All Comments” screen. The table has three columns (Rank, Feature Request, and Explanation) and eight rows, with features sorted in order of importance based on their Wave 3 ranking. The table summarizes which moderation features participants prioritized and how each feature is intended to work.
    The eight highest-ranked features are: (1) Customizing the Response to Detected Hate Speech, which allows users to choose between removing, blurring, hiding behind a content warning, or moving a comment to a “review later” folder; (2) Filtering Based on Words, which removes comments containing a user-specified word or string; (3) Changing Filter Decisions, which enables users to review and revise filtering decisions; (4) Hiding Comment Sections, which allows users to toggle the visibility of entire comment sections; (5) Filtering Based on Tone, which removes comments with an aggressive, insulting, or condescending tone; (6) Filtering Based on Topic, which removes comments related to user-specified topics; (7) Filtering Based on Users, which removes comments from selected users or groups; and (8) Filtering Based on Communication Pattern, which removes comments that match specific communication patterns.}
    \caption{Feature requests for the screen \textit{All Comments}, based on input from 40 activist social media users who experienced hate speech directed at them. The table provides the final ranking from Wave 3, the name of the feature, and an explanation of how the feature works. Requests are ordered by their Wave 3 rank.}
    \begin{tabular}{l l p{9cm}}
    \toprule
    \textbf{Rank} & \textbf{Feature Requests \underline{\textit{All Comments}}} & \textbf{Explanation} \\
    \midrule
     1. & Customizing the Response to Detected Hate Speech & Choose between removing, blurring, hiding behind a content warning, or moving to a ``review later'' folder. \\
     2. & Filtering Based on Words & Remove \textit{all comments} that contain a user-specified string. \\
     3. & Changing Filter Decisions & Possibility to review and revise filtering decisions if necessary. \\
     4. & Hiding Comments Sections & Possibility to toggle the visibility of all comment sections on or off. \\
     5. & Filtering Based on Tone & Remove \textit{all comments} with aggressive, insulting, or condescending tone.\\
     6. & Filtering Based on Topic & Remove \textit{all comments} that relate to topics specified by the user.\\
     7. & Filtering Based on Users & Remove \textit{all comments} from user-specified users or groups of users.\\
     8. & Filtering Based on Communication Pattern & Remove \textit{all comments} matching specific communication patterns.\\
    \bottomrule
    \end{tabular}
\label{tab:AllCommentsFeaturesExplanation}
\end{table*}

This section presents the personal moderation features that participants requested to explicitly apply to the screen \textit{All Comments} (see Figure \ref{fig:AllComments} for a visual representation of this screen and Table \ref{tab:AllCommentsFeaturesExplanation} for an explanation of the features).
All feature requests for the screen \textit{All Comments}, their associated ratings in Wave~1 and Wave~2, and their final ranking in Wave~3 can be found in the Appendix (Table~\ref{tab:AllCommentsFeatures}).

For the screen \textit{All Comments}, content-based features that allow for oversight and reversibility are both rated and ranked highly. For example, the feature \textit{Customizing the Response to Detected Hate Speech} ranked first (93.9\% W1, 92.3\% W2) (see Table \ref{tab:AllCommentsFeaturesExplanation} for feature explanations). Another feature request allowing for reversibility is the feature \textit{Changing Filtering Decisions}, which was ranked third. More than nine in ten participants rated this feature as important (added in W2, 92.3\% W2). 

Content-based personal moderation features that rely on user input are also frequently requested for the screen \textit{All Comments}. For instance, the second most important feature for the screen \textit{All Comments} was \textit{Filtering Based on Words}. More than nine in ten participants rated it as important (90.9\% W1, 94.9\% W2). Another feature request based on user input is the feature request \textit{Filtering Based on Topics}, which was ranked sixth (66.7\% W1, 92.3\% W2). 

Furthermore, content-based features that rely on advanced hate speech detection mechanisms are requested but ranked lower, e.g. \textit{Filtering Based on Tone} ranked fifth. More than seven in ten participants rated this feature as important in Wave~2 (71.8\%). Similarly, the feature request \textit{Filtering Based on Communication-Patterns} ranked last in Wave~3. This suggestion was seen as less important, with only slightly more than five in ten participants rating it as important (added in W2, 56.4\% W2).  

Finally, account-based personal moderation features were requested: The seventh most important feature in Wave~3 was \textit{Filtering Based on Users} (added in W2, 82.1\% W2). 

Overall, these findings highlight the importance of oversight and reversibility and input-driven features for personal moderation of \textit{All Comments}.

\subsubsection{Screen: Comments Own Posts}

\begin{table*}[t!]
\Description{Table 4 presents the final Wave 3 ranking of feature requests for the “Comments Own Posts” screen. The table contains three columns (Rank, Feature Request, and Explanation) and nine rows, with features ordered by importance. The table summarizes the specific moderation capabilities participants wanted for managing comments on their own posts.
The nine highest-ranked features are: (1) Customizing the Response to Detected Hate Speech, which allows users to choose between removing, blurring, hiding behind a content warning, or moving a comment to a “review later” folder; (2) Filtering Based on Users, which removes comments from selected users or groups of users; (3) Changing Filtering Decisions, which enables users to review and revise filtering decisions; (4) Filtering Based on Words, which removes comments containing a user-specified word or string; (5) Blocking of Replies to Selected Comments, which automatically prevents replies to user-selected comments; (6) Manually Allowing Comments, which lets users review and approve comments before they appear; (7) Post-Based Filtering Settings, which allow users to define filtering presets for individual posts; (8) Filtering Based on Topics, which removes comments related to user-specified topics; and (9) Listing Trusted Commenters, which allows only approved users to comment freely while others are filtered.}
\caption{All feature requests for the screen \textit{Comments Own Posts}, based on input from 40 activist social media users who experienced hate speech directed at them. The table provides the final ranking from Wave 3, the name of the feature, and an explanation of how the feature works. Requests are ordered by their Wave 3 rank.}
\label{tab:CommentsOwnPostsExplanation}
    \begin{tabular}{l l p{9cm}}
    \toprule
    \textbf{Rank} & \textbf{Feature Request \underline{\textit{Comments Own Posts}}} & \textbf{Explanation}\\
    \midrule
    1. & Customizing the Response to Detected Hate Speech & Choose between removing, blurring, hiding behind a content warning, or moving to a ``review later'' folder.\\
    2. & Filtering Based on Users & Remove \textit{comments on own posts} from user-specified users or groups of users.\\
    3. & Changing Filtering Decisions & Possibility to review and revise filtering decisions if necessary.\\
    4. & Filtering Based on Words & Remove \textit{comments on own posts} containing a user-specified string.\\
    5. & Blocking of Replies to Selected Comments & Automatically block replies to \textit{specific comments on own posts} that are manually selected by the user.\\
    6. & Manually Allowing Comments & Allow users to review and approve \textit{comments} before they appear \textit{under their posts}.\\
    7. & Post-Based Filtering Settings & Enable users to set specific filtering presets for \textit{individual own posts} before publishing.\\
    8. & Filtering Based on Topics & Remove \textit{all comments on their own posts} that relate to topics specified by the user.\\
    9. & Listing Trusted Commenters & Allow only approved users or followers to \textit{comment on their posts} without restrictions, while others are filtered.\\
    \bottomrule
    \end{tabular}
\end{table*}

This section outlines the personal moderation features that participants associated with the screen \textit{Comments on Own Posts} (see Figure \ref{fig:owncomments} for a visual representation of this screen and Table \ref{tab:CommentsOwnPostsExplanation} for an explanation of the features).

Content-based features that enable oversight and reversibility were ranked highest for the screen \textit{Comments Own Posts}. For example, \textit{Customizing the Response to Detected Hate Speech} was ranked first by the participants in Wave~3 (94.6\% W1, 94.7\% W2). Further, the possibility to \textit{Change Filtering Decisions} was requested by the participants via free text input in Wave~2. This feature ranked third in Wave~3 and was rated as important by more than nine in ten participants (92.1\%).

Moreover, account-specific features were requested with differences in their ranking: \textit{Filtering Based on Users}, (added in W2, 81.6\% W2) was ranked second, while the feature \textit{Listing Trusted Commenters} (added in W3) ranked ninth.

Several input-based features were requested for the screen \textit{Comments Own Post.} The fourth most important feature request was \textit{Filtering Based on Words}. More than eight in ten participants rated it as important in Wave~1 (89.2\%) and more than nine in ten in Wave~2~(93.3\%). Participants also suggested the \textit{Automated Blocking of Replies to Selected Comments}, which ranked fifth (added in W2, 76.3\% W2) or \textit{Manual Approving Comments}, which ranked sixth (added in W2, 71.1\% W2). Further, participants also suggested \textit{Post-Based Filtering Settings}, which ranked seventh (76.3\% W2). P31 explained this feature request as the possibility to ``\textit{choose a filtering preset when publishing a post}'' that will then apply to all comments on that post. Lastly, the feature \textit{Filtering Based on Topics} ranked eighth (W1 64.9\%, W2 81.6\%).

Features that allow for customized responses, user-based blocking, and revising filtering decisions ranked highest for personal moderation features on the screen \textit{Comments Own Post}. This indicates that participants prioritized flexible control mechanisms for managing comments on their own posts.

\subsubsection{Screen: Reels Tab}
\begin{table*}[t!]
\Description{Table 5 presents the final Wave 3 ranking of feature requests for the “Reels Tab” screen, with features ordered by importance. The table has three columns (Rank, Feature Request, and Explanation) and nine rows. It describes the moderation and control features participants wanted when interacting with Instagram Reels.
The nine features are: (1) Customizing the Response to Detected Hate Speech, which allows users to remove, blur, hide behind a content warning, or move reels to a “review later” folder; (2) Manual Video Controls, which let users control when reels play and display topic labels, headlines, or trigger warnings before playback; (3) Safe Space Mode, which provides a pre-configured setting with strong filtering; (4) Filtering Based on Topics, which removes reels related to user-specified topics; (5) Filtering Based on Words, which removes reels containing a user-specified string; (6) Filtering Based on Audio Content, which mutes or removes reels containing selected audio elements; (7) Filtering Based on Users, which removes reels from user-specified individuals or groups; (8) Filtering Based on Comments on Reel, which removes reels whose comment sections contain aggressive or hateful sentiment; and (9) Filtering Based on Format of Reel, which removes reels based on format characteristics, such as live or stitched videos.}
\caption{All feature requests for the screen \textit{Reels Tab}, based on input from 40 activist social media users targeted by hate speech. The table provides the final ranking from Wave 3, the name of the feature, and an explanation of how the feature works. Requests are ordered by their Wave 3 rank.}
    \begin{tabular}{l l p{9cm}}
    \toprule
    \textbf{Rank} & \textbf{Feature Requests \underline{\textit{Reels Tab}}} & \textbf{Explanation} \\
    \midrule
    1. & Customizing the Response to Detected Hate Speech & Choose between removing, blurring, hiding behind a content warning, or moving to a ``review later'' folder.\\
    2. & Manual Video Controls & Allow users to control when \textit{reels} play and display topic, headline, and trigger warnings before playing the reel.\\
    3. & Safe Space Mode & Provide a pre-configured mode with ``maximum filtering'' (P8) \textit{for reels}.\\
    4. & Filtering Based on Topics & Remove \textit{all reels} that relate to topics specified by the user.\\
    5. & Filtering Based on Words & Remove \textit{all reels} containing a user-specified string.\\
    6. & Filtering Based on Audio Content & Mute or remove \textit{all reels} containing user-specified audio elements.\\
    7. & Filtering Based on Users & Remove \textit{all reels} from user-specified users or groups of users.\\
    8. & Filtering Based on Comments on Reel & Remove \textit{reels} if their comments contain aggressive or hateful sentiment.\\
    9. & Filtering Based on \textit{Format of Reel} & Remove \textit{reels} based on their format, such as live or stitched videos.\\
    \bottomrule
    \end{tabular}
    \label{tab:ReelsTabExplanation}
\end{table*}

This section outlines the personal moderation features that participants associated with the screen \textit{Reels Tab} (see Figure \ref{fig:reels} for a visual representation of this screen and Table \ref{tab:ReelsTabExplanation} for an explanation of the features).

For the screen \textit{Reels Tab}, features allowing for reversibility and oversight ranked the highest. \textit{Customizing the Response to Detected Hate Speech} was ranked as the most important feature (81.8\% W1, 91.9\% W2) and the second-ranked feature was \textit{Manual Video Controls} (added in W2, 88.9\%). P31 further specified this feature request by emphasizing the importance of showing \textit{``topic, headline, and a trigger warning [...] across the entire screen''}, in addition to the video controls. This would allow for an informed decision of whether to watch the Reel or not (P25). 

Participants also suggested features that require a high level of automation. For example, participants requested a \textit{Safe Space Mode}, which ranked third (added in W3). This feature envisions a pre-configured environment with ``\textit{maximized filtering settings}'' (P8). \textit{Filtering Based on Comments on Reel} was also suggested for the \textit{Reels Tab} and ranked eighth (added in W3). This personal moderation feature presents a \textit{``content warning [encompassing the entire screen] if the comments under a Reel contain aggressive or hateful sentiment—even before playing the video''} (P31).  

Two input-based features were requested: \textit{Filtering Based on Topics} and \textit{Filtering Based on Words}. The former was ranked as the fourth most important feature. Almost eight in ten participants (78.8\%) rated this feature as important in Wave~1, and nearly all participants did so in Wave~2 (94.6\%). The latter ranked fifth in Wave~3. Almost seven in ten participants rated it as important in Wave~1 (69.7\%), and support rose to over nine in ten in Wave~2 (91.9\%).

Further content-based feature requests extended filtering to different content types. \textit{Filtering Based on Audio Content} ranked sixth (added in W2, 85.7\% W2). \textit{Filtering Based on Format of the Reel} ranked ninth (added in W2, 74.3\% W2) and describes filtering by video style or production type (e.g., live streams or stitched videos). Lastly, an account-specific feature was requested, but it ranked lower: \textit{Filtering Based on Users} ranked seventh (added in W2, 75.7\% W2).

Overall, personal moderation features enabling customization of system responses, manual video controls, and a ``safe space'' environment were ranked highest for the screen \textit{Reels Tab}.

\subsubsection{Screen: Home Feed}
\begin{table*}[t!]
\Description{Table 6 presents the final Wave 3 ranking of feature requests for the “Home Feed” screen, with features ordered by importance. The table has three columns (Rank, Feature Request, and Explanation) and ten rows. It describes the types of personal moderation features participants wanted when viewing their Home Feed.
The ten features are: (1) Differentiating Between Hate and Counter Speech, which reliably distinguishes harmful hate speech from legitimate counter speech; (2) Customizing the Response to Detected Hate Speech, which allows users to remove, blur, hide behind a content warning, or move content to a “review later” folder; (3) Filtering Based on Words, which removes content containing a user-specified string; (4) Filtering Based on Users, which removes content generated by selected users or groups; (5) Context-Sensitive Filtering, which adapts filtering decisions based on contextual factors like time or location; (6) Filtering Based on Topics, which removes content related to user-specified topics; (7) Filtering Based on Image Content, which removes content containing specified image elements; (8) Filtering Based on Video Content, which removes content containing specified video elements; (9) Filtering Based on Audio Content, which removes content containing specified audio elements; and (10) Filtering Based on Examples, which allows users to train the system by providing sample content to refine filtering behavior.}
\caption{All feature requests for the screen \textit{Home Feed}, based on input from 40 activist social media users who experienced hate speech directed at them. The table provides the final ranking from Wave 3, the name of the feature, and an explanation of how the feature works. Requests are ordered by their Wave 3 rank.}
    \begin{tabular}{l l p{9cm}}
    \toprule
    \textbf{Rank} & \textbf{Feature Request \underline{\textit{Home Feed}}} & \textbf{{Explanation}}\\
    \midrule
    1. & Differentiating Between Hate and Counter Speech & Reliably distinguish between harmful hate speech and legitimate counter speech on the Home Feed.\\
    2. & Customizing the Response to Detected Hate Speech & Choose between removing, blurring, hiding behind a content warning, or moving to a ``review later'' folder.\\
    3. & Filtering Based on Words  & Remove \textit{all content from Home Feed} containing a user-specified string.\\
    4. & Filtering Based on Users  & Remove \textit{all content from Home Feed} that is generated by specific users or groups of users.\\
    5. & Context-Sensitive Filtering & Automatically adapt filtering decisions based on contextual factors such as time or location.\\
    6. & Filtering Based on Topics  & Remove \textit{all content from Home Feed} that relates to topics specified by the user.\\
    7. & Filtering Based on Image Content  &  Remove \textit{all content from Home Feed} containing user-specified image elements.\\
    8. & Filtering Based on Video Content  &  Remove \textit{all content from Home Feed} containing user-specified video elements.\\
    9. & Filtering Based on Audio Content  &  Remove \textit{all content from Home Feed} containing user-specified audio elements.\\
    10. & Filtering Based on Examples  & Allow users to train the system by providing sample content to refine filtering behavior.\\
    \bottomrule
    \end{tabular}
\label{tab:HomeFeedExplanation}
\end{table*}

This section presents the personal moderation features that participants requested for the screen \textit{Home Feed} (see Figure \ref{fig:homefeed} for a visual representation of this screen and Table \ref{tab:HomeFeedExplanation} for an explanation of the features).

For the screen \textit{Home Feed}, content-based features that rely strongly on automation and require minimal user configuration were requested. For example, \textit{Differentiating Between Hate and Counter Speech} ranked first (added in W2, 91.2\% W2). P13 gives the example of ``\textit{a post [that] is clearly anti-Semitic or merely discussing anti-Semitism.}'' P14 said: ``\textit{Why is that person allowed to hate, but I'm not allowed to defend people?}''. Additionally, \textit{Context-Sensitive Filtering} was suggested and ranked fifth (added in W2, 67.7\% W2). 

The only suggested feature allowing for reversibility and oversight for the screen \textit{Home Feed} is the  content-based feature \textit{Customizing the Response to Detected Hate Speech} which ranked second (96.3\% W1, 94.1\% W2). 

Several input-driven features were requested, such as \textit{Filtering Based on Words}, which ranked third~(added in W2, 97.1\% W2~). Further, \textit{Filtering Based on Topics} (81.5\% W2, 91.2\% W2, ranked sixth W3) and \textit{Example-based Filtering} (added in W2, 67\% in W2, ranked tenth W3), were requested.

One account-specific feature was suggested, namely, \textit{Filtering Based on Users}. It ranked fourth in Wave~3 (added in W2, 88.2\%). Participants also suggested filters based on specific content types, though these were ranked lower. \textit{Filtering Based on Image Content} ranked seventh (added in W2, 76.5\% W2). \textit{Filtering Based on Video Content} ranked eighth (added in W2, 73.5\% in W2). Similarly, \textit{Filtering Based on Audio Content} ranked ninth (added in W2, 67.7\% in W2). 

Overall, participants highlighted the importance of nuanced filtering (e.g., distinguishing between hate speech and counter speech), customization options regarding how hate speech is handled, and reliable keyword or user-based blocking, while giving lower priority to modality-specific features.

\subsubsection{Screen: All DMs}
\begin{table*}[t!]
\Description{Table 7 presents the final Wave 3 ranking of feature requests for the “All DMs” screen, with features ordered by importance. The table has three columns (Rank, Feature Request, and Explanation) and eight rows. It summarizes the moderation and filtering capabilities participants wanted for managing all direct messages.The eight features are: (1) Customizing the Response to Detected Hate Speech, which allows users to remove, blur, hide behind a content warning, or move DMs to a “review later” folder; (2) Differentiating Between Hate and Counter Speech, which distinguishes hateful messages from legitimate counter speech or educational discussion; (3) Filtering Based on Communication Patterns, which removes DMs showing specific communication patterns; (4) Filtering Based on Users, which removes DMs from selected users or groups; (5) Filtering Based on Words, which removes DMs containing a user-specified string; (6) Filtering Based on Image Content, which removes DMs containing user-specified image elements; (7) Filtering Links, which allows users to block all DMs that contain links; and (8) Filtering Based on Language, which automatically adapts filtering based on the language of each DM.}
\caption{All feature requests for the screen \textit{All DMs}, based on input from 40 activist social media users targeted by hate speech. The table provides the final ranking from Wave 3, the name of the feature, and an explanation of how the feature works. Requests are ordered by their Wave 3 rank.}
    \begin{tabular}{l l p{9cm}}
    \toprule
    \textbf{Rank} & \textbf{Feature Request \underline{\textit{All DMs}}} & \textbf{Explanation}\\
    \midrule
    1. & Customizing the Response to Detected Hate Speech & Choose between removing, blurring, hiding behind a content warning or moving to a ``review later'' folder.\\
    2. & Differentiating Between Hate and Counter Speech & Reliably distinguish between \textit{hateful DMs} and legitimate counter speech or educational discussion.\\
    3. & Filtering Based on Communication-Patterns & Remove \textit{all DMs} exhibiting specific communication patterns.\\
    4. & Filtering Based on Users & Remove \textit{all DMs} from user-specified users or groups of users.\\
    5. & Filtering Based on Words & Remove \textit{all DMs} containing a user-specified string.\\
    6. & Filtering Based on Image Content &  Remove \textit{all DMs} containing user-specified image elements.\\
    7. & Filtering Links & Possibility block \textit{all DMs} that contain links.\\
    8. & Filtering Based on Language & Automatic adapt filtering based on the language of the \textit{respective DM}.\\
    \bottomrule
    \end{tabular}
    \label{tab:AllDMsExplanation}
\end{table*}

This section outlines the personal moderation features that participants associated with the screen \textit{All DMs} (see Figure \ref{fig:alldms} for a visual representation of this screen and Table \ref{tab:AllDMsExplanation} for an explanation of the features).

As in other conversational spaces, participants requested content-based features that enable oversight and reversibility for a personal moderation tool. For the screen \textit{All DMs}, participants ranked \textit{Customizing the Response to Detected Hate Speech} first, and it received high importance ratings across both waves (96.9\% W1, 97.1\% W2).

Participants also requested content-based features that strongly rely on automation and require minimal user configuration. For example, it is requested that a personal moderation tool can reliably \textit{Differentiate Between Hate and Counter Speech}. This feature ranked second (added in W2, 91.2\% W2). This request reflects a critical concern: while participants want effective filtering of harmful content, they also want to ensure that valuable counter speech—such as responses that challenge, reframe, or call out hate—remains visible. Another example is the request to \textit{Filter Based on Communication Patterns}, which ranked third in Wave~3 (added in W2, 76.5\% W2). Participants saw value in detecting repetitive harassment patterns, such as coordinated targeting or persistent unwanted contact.

One account-specific feature was requested: \textit{Filtering Based on Users}. It was ranked the fourth most important feature (added in W2, 85.3\% W2). The input-driven feature \textit{Filtering Based on Words} ranked fifth and gained importance across waves (78.1\% W1, 94.1\% W2). 

Content type and format-specific options, namely \textit{Filtering Based on Image Content} (added in W3, ranked 6th W3), \textit{Filtering Links} (added in W2, 73.5\%, ranked 7th W3), and \textit{Filtering Based on Language} (added in W3, ranked 8th W3), were ranked lowest. These options were perceived as less important for the screen \textit{All DMs}.

Overall, participants emphasized flexible response options, accurate distinction between hate and counter speech, and filters based on communication patterns as the most important tools for the personal moderation of direct messages.

\section{Discussion}

We showed where users want personal moderation (RQ1) and which features they value most (RQ2). In this discussion, we situate these findings within existing literature and highlight how our results offer a conceptualization for the granularity of control in personal moderation. Further, we show how requested features can be categorized and related to different screens. Finally, we provide design recommendations for personal moderation tools on social media.

\subsection{The Where of Personal Moderation}

\subsubsection{Prioritizing Screens for Effective Personal Moderation}

Our findings highlight that the need for personal moderation differs across screens. By prioritizing which screens users consider most critical, we provide actionable guidance for researchers exploring personal moderation, designers developing user interfaces, and platforms allocating resources. This prioritization reduces implementation complexity and ensures that efforts target the contexts where personal moderation has the greatest impact. 

Another contribution is identifying screens as a meaningful unit of control for personal moderation, an issue raised by previous work~\cite{Jhaver2023}. The variations in importance distributions across screens, combined with high ratings of a subset of them, show the merit of this conceptualization. This granularity is crucial because meaningful design guidance requires understanding not only which features users value but also where they expect to apply them.
Future research should dive deeper into the ``screen'' as a relevant axis for personal moderation. Accordingly, platforms should modify personal moderation features so that they can be applied separately to different screens. Offering screen-specific settings would extend existing features while aligning with user priorities as established in our study. However, the identified screens and their associated set of affordances used in our study are specific to Instagram~\cite{arslan2021towards, Ronzhyn2023Affordances}. Further research is necessary to establish the preferences for screen-based personal moderation on other platforms.

\subsubsection{Labor \& Usability as Design Constraints for Personal Moderation Tools}

Designing a personal moderation tool that takes into account the screen-specific differences identified in our study raises new issues regarding usability and labor. Personal moderation tools must be designed so they remain easily accessible and are actually used by those who need them. This tension is familiar from existing features such as the Home Feed on Instagram, which has the ``For You'' Tab (algorithmic curation of content) and the ``Following'' Tab (chronological feed of followed accounts). Despite the explicit wish for a chronological feed as documented in previous work~\cite{Rogers2022User, Luther2025}, users are often unaware of this possibility~\cite{Luther2025}. This issue is likely even exacerbated when designing a personal moderation tool encompassing all features for all screens discussed in this work, as this challenges implementation to reflect an even larger complexity. To address this challenge, we highlight the value of the ranking provided by this work. This allows for a focus on the most highly ranked screens and features, which helps developers and researchers prioritize their efforts. This prioritization further has the potential to reduce user burden when configuring these tools, an issue frequently identified by previous work on personal moderation~\cite{Jhaver2023, Heung2025Ignorance}. 

Our participants highlighted the labor involved in configuring personal moderation settings when asked about low-rated screens, while this was not mentioned for high-rated screens. This could suggest that while participants are aware of the labor involved, they are willing to put in the effort for the screens where they deem personal moderation as highly impactful. Nonetheless, configuring personal moderation for a subset of screens still necessitates both an awareness and understanding of the personal moderation possibilities and a (at least initial) time investment. Both lacking awareness and lacking the necessary time resources have been previously identified as hindrances in adoption, e.g., of privacy-related features~\cite{charnsethikul2025navigating}. Together, these considerations underscore that labor is a central design constraint for personal moderation tools. Thus, future work should explore how to balance controllability with usability when configuring personal moderation. While high controllability empowers users to tailor their moderation, excessive configuration effort risks creating cognitive burdens~\cite{Jhaver2023}. We recommend designing interfaces where personal moderation tools are visible and discoverable, yet non-intrusive for users less affected by hate speech. Research should investigate strategies for progressive disclosure as a potential pathway to address this. Further, providing meaningful defaults is central as well as designing personal moderation tools in a way that ensures a shared understanding of its functionalities, issues highlighted by previous work~\cite{Jhaver2023}.

\subsubsection{Personal Moderation on Conversational Screens}
Our findings underscore the high importance of personal moderation in conversational spaces such as comment sections and direct messages. Prior work has shown that conversational settings are challenging for moderation, as meaning is context-dependent, norms vary across communities, and interpersonal histories shape interpretation~\cite{Seering2019Moderator, Dorn2024HarmfulSpeech}. Our study shows that for users, these spaces are precisely where personal moderation features are most needed, as users often know which words, topics, or individuals are likely to cause harm. This highlights a unique role for personal moderation: rather than competing with automated or platform-wide systems, it can give users control in the interpersonal contexts where automated hate speech detection suffers from limitations~\cite{hartmann2025lost}. 
As these insights are grounded in Instagram’s conversational spaces, generalizability to other platforms is hypothetical. We do, however, assume, that platforms that are structured similarly may benefit equally. Messaging, comment threading, and moderation practices differ across platforms, as do community norms, which can shape both exposure to harmful content and the perceived usefulness of personal moderation. 
Future work should examine how personal moderation needs in conversational contexts translate to other platforms.

\subsubsection{Personal Moderation on Algorithmically Curated Screens}
Our results show that personal moderation is also considered important when it comes to algorithmically curated screens such as the \textit{Home Feed} and the \textit{Reels Tab}. This finding aligns with critiques of algorithmic opacity and power asymmetries in platform governance~\cite{gillespie2018custodians, eslami2024}. Research surrounding user control in algorithmically curated spaces has, however, mostly looked at that issue from the perspective of improving recommendation systems~\cite{Harambam2019RecSys, Stanislaw2025RecSys} rather than addressing it from a moderation perspective. 

Yet, moderation and curation are deeply intertwined: when users do not wish to see certain content, the distinction between filtering out and just not recommending becomes blurry. What does not get recommended can effectively be experienced as “moderated out”~\cite{Jhaver2023}. This is particularly relevant given that recommendation is not a neutral process, but one that is shaped by commercial logics which prioritize minimizing advertiser risk over addressing user harms~\cite{milano2020recommender}. Our findings suggest that personal moderation could rebalance this asymmetry by giving users greater agency to define not only what is highlighted to them, but also what should be suppressed, thereby complementing platform-driven curation with personal moderation. However, our findings are also shaped by Instagram’s screen architecture and its associated affordances~\cite{arslan2021towards, Ronzhyn2023Affordances}. Other platforms vary in feed design, ranking algorithms, and transparency, which may influence the utility of personal moderation on algorithmically curated screens on other platforms. Further research is needed to identify which of the generated insights generalize and which are platform-specific.

\subsection{The What of Personal Moderation}
In the following, we will categorize the five most highly ranked feature requests of all the discussed screens. Then we show how this categorization relates to the screens for which they were requested.

\subsubsection{Oversight and Reversibility as Core Pillars of Personal Moderation}
The category \textit{Oversight \& Reversibility Features} is based on previous work and describes content-based features that enable users to supervise moderation outcomes and to adjust them to their needs~\cite{Heung2025Ignorance}. Oversight and reversibility features requested in our study include \textit{Customizing the Response to Detected Hate Speech, Changing Filtering Decisions} and \textit{Manually Allowing Comments}. 

Among these, \textit{Customizing the Response to Detected Hate Speech} emerged as especially important in our study. Its importance suggests that this feature could address previously identified issues surrounding personal moderation. One common issue is the fear of missing out on content~\cite{Heung2025Ignorance, jhaver2025personal}. By allowing users to review the filtering decisions of a personal moderation tool, users do not have to fear that they will miss out on content, as they can always access all content in a ``review'' folder. Also, the limited effectiveness of content and trigger warnings due to the limited customization to individual needs, contexts, or vulnerabilities~\cite{Charles2022Typology, willems2025trigger} could be addressed with such a feature.

Most oversight and reversibility features were suggested for the screens \textit{All Comments} and \textit{Comments Own Post}. The experiences of our participants with malfunctioning moderation systems on these screens could explain this increased need to be able to control and, if necessary, reverse decisions of a personal moderation tool. 

We, therefore, recommend the implementation of content-based oversight and reversibility features within a personal moderation tool. Designers should explore response customization, such as choosing between removing, blurring, content warnings, and moving content to a separate folder for later review. Combining this with the ability to change \added{personal} moderation decisions could further mitigate frustration and strengthen trust~\cite{Heung2025Ignorance}.

Overall, our findings highlight oversight and reversibility as a core pillar of effective personal moderation across screens, with particular high importance in moderating comment sections. 

\subsubsection{Input-driven Features Are Most Important Where the Details of Hate Speech Are Known}

We define input-driven features as personal moderation features that rely on the input of the user to explicitly specify which kinds of content they want to moderate. Thus, this can also be categorized as content-based personal moderation according to ~\citeauthor{Jhaver2023} (\citeyear{Jhaver2023}). In our study, \textit{Filtering Based on Words}, \textit{Filtering Based on Topics}, and \textit{Filtering Based on Examples} emerged as input-driven features. 

Word filters, in particular, are already integrated into several social media platforms. For example, Instagram allows word filters for comments and message requests~\cite{instagram_privacy_settings} and Facebook enables word filters for comments, including options for linguistic variants or bulk upload~\cite{facebook_hate_speech_help}. Our findings, however, show that it is important for users to distinguish which words are filtered on which screen. To filter the same words for all content on the platform does not fulfill the user needs we identified. This can potentially address the previously highlighted shortcomings of word filters such as context insensitivity~\cite{Jhaver2023}. Topic filters have likewise been explored in prior work~\cite{bhargava2019gobo, salminen_taking_2021}, but platform implementations tend to focus on surfacing desired content rather than avoiding harmful or unwanted topics~\cite{bhargava2019gobo, Stanislaw2025RecSys, Harambam2019RecSys}. Our work further emphasizes the importance of the moderation perspective for input-driven features highlighted by previous work~\cite{Jhaver2023}.  

Input-driven features that apply to the comment sections (both \textit{All Comments} and \textit{Comments on Own Posts}) were ranked particularly high by our participants. The heightened importance of input-driven features of comments suggests that users see these features as especially effective on screens where hate speech is both frequent and predictable. In contrast to that, input-driven features were deemed less important for algorithmically curated screens such as the \textit{Home Feed} or \textit{Reels Tab}. A potential reason is that, on these screens, users cannot easily anticipate what content will appear, making it more difficult to define effective filters based on words, topics, or examples. This differentiation between conversational and algorithmic screens adds a novel, screen-specific perspective to the design of input-driven features.

Taken together, input-driven features remain a foundational element of personal moderation. However, our results emphasize that their utility is strongest for conversational screens.

\subsubsection{Content-Type Specific Features Fill Critical Gaps in Algorithmic Feeds and DMs}

Content-type specific features describe features that adjust filtering based on the format of the content, such as text, images, or videos. Content-type specific features in our study include \textit{Filtering Based on Audio, Image, or Video Content, Filtering Based on Format of Reel}, and \textit{Filtering Links}.

In our study, content-type specific features were especially valued on algorithmically curated screens such as the \textit{Reels Tab} and \textit{Home Feed}. Participants highlighted that these feeds often expose them to unanticipated, multimedia content, making text-only filtering insufficient.

On \textit{All DMs}, a screen without algorithmic curation, participants requested image and link filters. This indicates that, beyond text, participants considered filters for other content types relevant in private interactions. Prior work has shown that personal moderation is often discussed in terms of text~\cite{Wang2025ExampleRuleLLM, salminen_taking_2021}, but our findings highlight that protecting users from hate speech conveyed through various content types is increasingly central.

Altogether, content-type specific features respond to a gap in existing tools by extending personal moderation to non-textual content and separating its filtering based on individual content types. We show that they are requested for algorithmically curated feeds to handle unpredictability, and for DMs to address targeted image- and link-based harassment.

\subsubsection{Account-Specific Features Are Essential but Underdeveloped}

We define account-specific features as features that allow users to shape their online environment around social ties, restricting or loosening restrictions of interactions with individual users or groups of users~\cite{jhaver2018online}. In our study, account-specific features include \textit{Filtering Based on Users} and \textit{Listing Trusted Commenters}. 

We find that account-specific features were ranked highly for three screens: \textit{Comments Own Posts}, \textit{Home Feed}, and \textit{All DMs}. For \textit{Comments Own Posts}, account-specific personal moderation is already possible on Instagram. Users can decide whether they want to allow comments from followers, people that one follows back, or everyone. It is also possible to dis- or enable whether people can comment on GIFs on your content. For the \textit{Home Feed}, one can toggle between the ``For You'' Tab (which is the default) and the ``Following'' Tab, which allows control over whose content one sees. For \textit{All DMs}, several control options are in place on Instagram~\cite{instagram_privacy_settings}. Since these options are already implemented on several platforms, this raises the question of which factors foster or hinder the adoption of these tools. Recent work showed the influence of psychosocial factors on the adoption of account-based personal moderation, namely the muting function on Facebook~\cite{jhaver2025personal}. This muting function allows users to remove the content of selected accounts from their news feed. The representative survey showed a significant negative influence of fear of missing out (FoMO) on users' intention to mute, while subjective norms and trust in Facebook moderation positively influenced the use of these tools~\cite{jhaver2025personal}.

This suggests two important insights for platform design. First, existing account-specific features are highly valued by users targeted by hate speech, but need to be extended to allow for more fine-grained control. Our rankings provide guidance on which features to prioritize for implementation. Second, as demonstrated by~\citeauthor{jhaver2025personal}~(\citeyear{jhaver2025personal}), the effectiveness of account-specific features in reducing exposure to harm extends beyond their technical implementation. Their adoption hinges on addressing psychosocial barriers, including fear of missing out (FoMO), and on cultivating trust in the platform’s moderation practices.

\subsubsection{Highly Automated Features Are Most Important Where The Details of Hate Speech Are Not Known}

Highly automated features describe content-based features that rely on advanced hate speech detection mechanisms and minimize manual user input or configuration. In our study, these features include \textit{Filtering Based on Tone, Filtering Based on Communication-Pattern, Safe Space Mode, Differentiating Between Hate and Counter Speech}, and \textit{Context-Sensitive Filtering}.

Highly automated features were requested on algorithmically curated screens (\textit{Reels Tab, Home Feed}), as they offered a way to reduce exposure to unexpected hate speech without requiring detailed manual configuration. For example, for the ~\textit{Reels Tab}, participants requested a \textit{Safe Space Mode}, which removed any distressing content. What content is distressing differs between participants. Thus, training a personal classifier on unwanted content provides a promising path. This work can be informed by the work of \citeauthor{Wang2025ExampleRuleLLM} (\citeyear{Wang2025ExampleRuleLLM}), which investigated the preferences of users for training a personalized content classifier. The results show the nuanced preferences of users between rule-based training, example-based training, and prompting an LLM~\cite{Wang2025ExampleRuleLLM}. Similar work on image, audio, and video data would be necessary to inform the design of personalized \textit{Safe Space Mode} for the \textit{Reels Tab}. 

Participants rarely requested such features for comment sections. This resonates with prior critiques of automated moderation, where ambiguity and evolving norms undermine trust~\cite{jhaver2022designing, Jhaver2023}. Our findings extend these critiques by showing that, while participants welcome automation on most screens, they remain wary of high automation when moderating \added{conversational screens}. 

\subsection{Personal Moderation in Context}

While our study examines the design space of personal moderation, situating these insights within broader literature highlights that personal moderation can only constitute one element of a larger socio-technical response to hate speech.

One critical concern is platform responsibility~\cite{Jhaver2023}. Increasing possibilities for personal moderation should not serve as a means to free platforms from their responsibility to remove content that undoubtedly merits platform-wide removal. Prior work has repeatedly highlighted the shortcomings of platform-wide moderation~\cite{Arshad-Ayaz2022Perspectives, jhaver2023users, Poletto2020Resources}, which also personal moderation functionalities can and should not be expected to address. The purpose of personal moderation should be to complement platform-wide moderation to address limitations that inevitably result from a one-size-fits-all approach to moderation. Meeting these platform-level responsibilities, however, depends on the expertise embedded in moderation processes. Previous work emphasized the expertise necessary in content moderation processes~\cite{Abdelkadir2025Expertise}. This stands in stark opposition to current industry trends of scaling back human moderation labor~\cite{meta_more_speech, eu_commission_rfi_x_2024, europarl_e002454_2024}. However, platform priorities are to limit additional financial investment into moderation efforts as much as possible, as long as it does not threaten investor interests or regulatory requirements~\cite{gayo-avello2015SoMe}.

A second central concern is that effective moderation cannot be separated from broader questions of accountability for those who spread hate speech. Prior work shows that consequences for people spreading hate speech online are inconsistent and often minimal~\cite{Siegel_2020}. This gap is multifaceted: it spans media literacy, platform governance, and legal structures. Measures such as suspensions warnings and community guidelines can serve as deterrents~\cite{felmlee2020can, Yildirim_Nagler_Bonneau_Tucker_2023}. Also democratic moderation approaches can be beneficial, yet they raise tensions around due process, transparency, and risks of majority-driven punitive action~\cite{chen2025democratic}. Complementary alternatives, including restorative justice approaches, offer potential pathways for addressing harm by centering repair rather than punishment~\cite{xiao2022sensemaking}. However, such models depend on sustained investment in trust and safety infrastructures, at a time when many platforms are reducing these resources ~\cite{eu_commission_rfi_x_2024, europarl_e002454_2024}.

There is also a need for a stronger legal framework to ensure consequences for offenders. Personal or platform moderation alone cannot address repeated and severe hate speech violations. Regulatory bodies can provide clarity by setting standards for escalation and ensuring that severe cases are pursued with adequate legal resources.

\subsection{Limitations \& Future Work}

While our study offers valuable insights, our research is shaped by contextual factors that introduce limitations that need acknowledgment. Our sample consisted of activists in Germany, which means that experiences of hate speech in other cultural or political contexts may differ. Nevertheless, focusing on this group allowed us to capture perspectives from individuals who are particularly exposed to hate speech and thus have well-informed needs regarding personal moderation. Furthermore, our sample is drawn from movements leaning towards the progressive side of the ideological spectrum. This may have influenced the perspectives and priorities expressed, since research points towards the fact that individuals who identify as liberal or left-leaning are more likely to view content moderation positively compared to conservatives~\cite{alizadeh2022content, shen2019discourse, Wang2023PerceptionOfContentModeration}. The changes in engagement across the three study waves might have introduced a bias by keeping the perspectives of users who are already in favor of personal moderation approaches. It should also be noted that we presented the screens in the same order across waves and participants. This could have introduced order effects, for example, due to fatigue, habituation, or increasing familiarity with the task over time. Since the amount of free-text input across screens remained relatively stable, there is no strong indication of decreasing engagement as the survey progressed; however, subtle effects on perceived importance or attentional focus cannot be ruled out. Similarly, the fixed order may have influenced participants’ relative ratings, as earlier screens could serve as anchors affecting judgments of later screens, or later screens might have benefited from learning effects in understanding moderation features. Finally, the screens we examined reflect Instagram’s current architecture. Although many major platforms organize user interactions into similar screens (e.g., feeds, messaging spaces, comment threads), the extent to which our findings transfer depends on many factors, e.g., how similarly these screens function in terms of features, interaction flows, and community norms. While our results should therefore be interpreted primarily in the context of Instagram, our screen-based abstraction provides a structured way to assess personal-moderation needs elsewhere, especially on platforms whose architectures share these core components.

Building on these limitations, the study opens up promising avenues for future research. Our results show that screens represent a meaningful and underexplored unit of analysis for personal moderation, and future work should establish this axis more firmly across different platforms. Particularly, work addressing the differences in personal moderation needs between conversational and algorithmically curated spaces provides a promising path for future research. Especially, researching and designing oversight and reversibility features promises to increase user agency and trust in personal moderation tools. In addition, future work should take a more holistic view by examining how platform, community, and personal moderation can be aligned to avoid redundancy or gaps. This is crucial for designing moderation ecosystems that balance user agency and safety. Together, these avenues point to a research agenda that treats personal moderation not as an isolated feature but as an integral component of the socio-technical system of online moderation.

\section{Conclusions}
In this paper, we examined how activist users targeted by hate speech envision personal moderation tools and how personal moderation needs differ across a platform’s diverse screens. Through a three-wave Delphi study, we iteratively distilled participants’ priorities, showing that the demand for personal moderation is highly screen-dependent, with conversational and algorithmically curated spaces requiring the strongest forms of user control. We further identified which feature types users value and how these preferences shift across screens.

Our work contributes to research and design by introducing screens as a conceptualization for granularity of control of personal moderation needs and by offering an empirically grounded prioritization of moderation features per screen. This screen-based perspective provides a practical and conceptual scaffold for building tools that align with the realities of everyday platform use.

Looking ahead, effective personal moderation must be situated within broader socio-technical responses to hate speech, including platform-level governance and accountability structures. This also requires investigating how personal moderation can be integrated and calibrated alongside platform and community moderation to form a coherent, multi-layered system. Future work should explore how screen-sensitive personal moderation tools perform in real-world deployment and across different platforms. By foregrounding how platform contexts shape user needs, this work provides an approach for developing personal moderation systems that align more closely with the complexities of contemporary online environments.

\begin{acks}
We thank all participants for their insights and the anonymous reviewers for their helpful, constructive feedback. This work is funded by the German Federal Ministry of Research, Technology and Space and the European Union - NextGenerationEU.
\end{acks}

\bibliographystyle{ACM-Reference-Format}
\bibliography{references}
\newpage
\appendix
\onecolumn
\section{Appendix}
\subsection{Importance Ratings of Screens across Wave~1, Wave~2 \& Wave~3}
\label{AllInteractionContexts}
\begin{table}[h!]
\Description{Table 8 provides an overview of the importance ratings of screens across three study waves: Wave 1 (W1), Wave 2 (W2), and Wave 3 (W3). The table has ten rows and four columns. The first column, Screen, lists nine screen types: All Comments, Comments Own Posts, Reels Tab, Home Feed, All DMs, Comments Specific Posts, DMs Specific Accounts, Explore Tab, and Search Results. The next three columns correspond to the three waves. For each screen and wave, two percentages are shown: the percentage of participants who did not rate the screen as important, indicated by a “cross” symbol, and the percentage of participants who did rate it as important, indicated by a “check” symbol.}
\caption{Overview of the Importance Ratings of Screens across W1, W2 \& W3. The screens are sorted by the share of participants who rate it as important in Wave~3.}
    \begin{tabular}{l l l l}
    \toprule
    \textbf{Screen} & \textbf{Importance W1} & \textbf{Importance W2} & \textbf{Importance W3}  \\
    \midrule
    All Comments &  \ding{52} 81.6\% \ding{56} 13.2\%  &  \ding{52} 92.1\%  \ding{56} 2.6\% &  \ding{52} 100.0\%  \ding{56} 0.0\% \\
    Comments Own Posts &   \ding{52} 84.6\%  \ding{56} 5.1\% &  \ding{52} 89.5\% \ding{56} 5.3\% &  \ding{52} 92.5\%  \ding{56} 2.5\% \\
    Reels Tab & \ding{52} 61.5\%  \ding{56} 15.4\%  &  \ding{52} 78.9\%  \ding{56} 7.9\% &  \ding{52} 87.5\%  \ding{56} 7.5\% \\
    Home Feed & \ding{52} 65.8\%  \ding{56} 28.9\%&  \ding{52} 76.3\%  \ding{56} 15.8\% &  \ding{52} 87.5\%  \ding{56}  10.0\%\\
    All DMs &  \ding{52} 71.8\%  \ding{56} 17.9\%  &  \ding{52} 81.6\% \ding{56} 15.8\%  &  \ding{52} 85.0\%  \ding{56} 7.5\% \\    
    Comments Specific Posts &  \ding{52} 72.5\%  \ding{56} 17.5\%  &  \ding{52} 76.3\%  \ding{56} 13.2\% &  \ding{52} 77.5\%  \ding{56} 7.5\% \\
    DMs Specific Accounts &  \ding{52} 57.5\% \ding{56} 35.0\%  & \ding{52} 73.7\% \ding{56} 10.5\% &  \ding{52}  75.0\% \ding{56} 12.5\% \\
    Explore Tab & \ding{52} 71.4\% \ding{56} 22.9\% &  \ding{52} 73.7\% \ding{56} 15.8\%  &  \ding{52} 72.5\% \ding{56} 15.0\%  \\
    Search Results & \ding{52} 41.1\%  \ding{56} 33.3\%  & - & - \\    
    \bottomrule
    \end{tabular}
    \label{tab:AllInteractionContexts}
\end{table}

\subsection{Importance Ratings (W1, W2) and Ranking (W3) of Features}
\begin{table}
\Description{Table 9 summarizes all feature requests for the screen All Comments, based on input from 40 activist social media users who experienced hate speech directed at them. The table has four columns and eight rows. The first column, Feature Requests, lists the requested features: Customizing the Response to Detected Hate Speech, Filtering Based on Words, Changing Filter Decisions, Hiding Comments Sections, Filtering Based on Tone, Filtering Based on Topic, Filtering Based on Users, and Filtering Based on Communication-Pattern.
The second column, Importance Wave 1, shows the percentage of participants who rated each feature as important, indicated by a check symbol, and the percentage who did not, indicated by a cross symbol. The third column, Importance Wave 2, uses the same format to show ratings from the second wave. The fourth column, Rank Wave 3, presents the final ranking of each feature in Wave 3 along with the percentage of total points it received.}
\caption{All feature requests for the screen \textit{All Comments}, based on input from 40 activist social media users who experienced hate speech directed at them. The table shows importance ratings from Wave~1 and Wave~2, as well as the final ranking and the percentage of points from Wave~3. Requests are ordered by their Wave~3 rank.}
    \begin{tabular}{l l l c}
    \toprule
    \textbf{Feature Requests \textit{All Comments}} & \textbf{Importance W1} & \textbf{Importance W2} & \textbf{Rank W3 (\% Points )}  \\
    \midrule
    Customizing the Response to Detected Hate Speech &\ding{52}  93.9\% \ding{56} 3.0\% & \ding{52}  92.3\% \ding{56} 0.0\% & 1. (13.8\%)\\
    Filtering Based on Words & \ding{52} 90.9\% \ding{56} 0.0\% & \ding{52} 94.9\% \ding{56} 2.6\% & 2. (12.2\%)\\
    Changing Filter Decisions & - &  \ding{52} 92.3\% \ding{56} 2.6\%  & 3. (11.0\%)\\
    Hiding Comments Sections & - & \ding{52} 76.9\% \ding{56} 10.3\% & 4. (10.0\%)\\
    Filtering Based on Tone & - &  \ding{52}  71.8\% \ding{56} 7.7\% & 5. (10.0\%)\\
    Filtering Based on Topic & \ding{52} 66.7\% \ding{56} 12.1\% & \ding{52}  92.3\% \ding{56} 0.0\% & 6. (9.0\%)\\
    Filtering Based on Users & - & \ding{52} 82.1\% \ding{56} 5.1\% & 7. (9.0\%)\\
    Filtering Based on Communication-Pattern& - &\ding{52}  56.4\% \ding{56} 25.6\% &8. (8.8\%)\\
    \bottomrule
\end{tabular}
\label{tab:AllCommentsFeatures}
\end{table}

\begin{table*}
\Description{Table 10 summarizes all feature requests for the screen Comments Own Posts, based on input from 40 activist social media users who experienced hate speech directed at them. The table has four columns and nine rows. The first column, Feature Requests, lists the requested features: Customizing the Response to Detected Hate Speech, Filtering Based on Users, Changing Filtering Decisions, Filtering Based on Words, Automated Blocking of Replies to Selected Comments, Manually Allowing Comments, Post-Based Filtering Settings, Filtering Based on Topics, and Listing Trusted Commenters. The second column, Importance Wave 1, shows the percentage of participants who rated each feature as important, indicated by a check symbol, and the percentage who did not, indicated by a cross symbol. The third column, Importance Wave 2, uses the same format to show ratings from the second wave. The fourth column, Rank Wave 3, presents the final ranking of each feature in Wave 3 along with the percentage of total points it received.}
\caption{All feature requests for the screen \textit{Comments Own Posts}, based on input from 40 activist social media users who experienced hate speech directed at them. The table shows importance ratings from Wave~1 and Wave~2, as well as the final ranking and the percentage of points from Wave~3. Requests are ordered by their Wave~3 rank.}
    \begin{tabular}{l l l c}
    \toprule
    \textbf{Feature Requests \textit{Comments Own Posts}} & \textbf{Importance W1} & \textbf{Importance W2} & \textbf{Rank W3 (\% Points)}\\
    \midrule
    Customizing the Response to Detected Hate Speech & \ding{52} 94.6\% \ding{56} 2.7\% &  \ding{52}  94.7\% \ding{56} 5.3\% & 1. (15.6\%)\\
    Filtering Based on Users & - &  \ding{52}  81.6\% \ding{56} 10.5\% & 2. (12.4\%)\\
    Changing Filtering Decisions & - & \ding{52}  92.1\% \ding{56} 7.9\% &3. (11.9\%)\\
    Filtering Based on Words & \ding{52} 89.2\% \ding{56} 2.7\% &  \ding{52} 93.3\% \ding{56} 6.7\% & 4. (11.7\%)\\
    Automated Blocking of Replies to Selected Comments & - &  \ding{52}  76.3\% \ding{56} 10.5\% & 5. (9.3\%) \\
    Manually Allowing Comments & - & \ding{52} 71.1\% \ding{56} 15.8\%&6. (9.1\%)\\
    Post-Based Filtering Settings & - &  \ding{52}  76.3\% \ding{56} 15.8\% & 7. (7.9\%)\\
    Filtering Based on Topics & \ding{52} 64.9\%  \ding{56}  21.6\%& \ding{52} 81.6\% \ding{56} 7.9\% &8. (7.7\%)\\
    Listing Trusted Commenters & - & - &9. (6.8\%)\\
    \bottomrule
\end{tabular}
\label{tab:CommentsOwnPosts}
\end{table*}

\begin{table}
\Description{Table 11 summarizes all feature requests for the screen Reels Tab, based on input from 40 activist social media users targeted by hate speech. The table has four columns and nine rows. The first column, Feature Requests, lists the requested features: Customizing the Response to Detected Hate Speech, Manual Video Controls, Safe Space Mode, Filtering Based on Topics, Filtering Based on Words, Filtering Based on Audio Content, Filtering Based on Users, Filtering Based on Comments on Reel, and Filtering Based on Format of Reel.
The second column, Importance Wave 1, shows the percentage of participants who rated each feature as important, indicated by a check symbol, and the percentage who did not, indicated by a cross symbol. The third column, Importance Wave 2, uses the same format to display ratings from the second wave. The fourth column, Rank Wave 3, presents the final ranking of each feature in Wave 3 along with the percentage of total points it received.}
\caption{All feature requests for the screen \textit{Reels Tab}, based on input from 40 activist social media users targeted by hate speech. The table shows importance ratings from Wave~1 and Wave~2, as well as the final ranking and the percentage of points from Wave~3. Requests are ordered by their Wave~3 rank.}
    \begin{tabular}{l l l c}
    \toprule
    \textbf{Feature Requests \textit{Reels Tab}} & \textbf{Importance W1} & \textbf{Importance W2} & \textbf{Rank W3 (\% Points )}\\
    \midrule
    Customizing the Response to Detected Hate Speech &\ding{52}  81.8\% \ding{56} 3.0\%  & \ding{52} 91.9\% \ding{56} 2.7\% & 1. (11.2\%)\\
    Manual Video Controls & - & \ding{52} 88.9\% \ding{56} 2.8\% & 2. (9.8\%)\\
    Safe Space Mode & - & - & 3. (9.3\%)\\
    Filtering Based on Topics&\ding{52}  78.8\% \ding{56} 0.0\% & \ding{52} 94.6\% \ding{56} 0.0\% & 4. (8.9\%)\\
    Filtering Based on Words & \ding{52} 69.7\% \ding{56} 9.1\%  & \ding{52} 91.9\% \ding{56} 2.7\%  & 5. (8.1\%)\\
    Filtering Based on Audio Content & - &\ding{52} 85.7\% \ding{56} 11.4\% & 6. (7.3\%)\\
    Filtering Based on Users & - &\ding{52} 75.7\% \ding{56} 2.7\% &7. (7.3\%)\\
    Filtering Based on Comments on Reel & - & - & 8. (6.6\%)\\
    Filtering Based on Format of Reel & - &\ding{52} 74.3\% \ding{56} 5.7\% & 9. (5.6\%)\\
    \bottomrule
\end{tabular}
\label{tab:ReelsTab}
\end{table}

\begin{table}
\Description{Table 12 summarizes all feature requests for the screen Home Feed, based on input from 40 activist social media users who experienced hate speech directed at them. The table has four columns and ten rows. The first column, Feature Requests, lists the requested features: Differentiating Between Hate and Counter Speech, Customizing the Response to Detected Hate Speech, Filtering Based on Words, Filtering Based on Users, Context-Sensitive Filtering, Filtering Based on Topics, Filtering Based on Image Content, Filtering Based on Video Content, Filtering Based on Audio Content, and Filtering Based on Examples.
The second column, Importance Wave 1, shows the percentage of participants who rated each feature as important, indicated by a check symbol, and the percentage who did not, indicated by a cross symbol. The third column, Importance Wave 2, uses the same format to show ratings from the second wave. The fourth column, Rank Wave 3, presents the final ranking of each feature in Wave 3 along with the percentage of total points it received.}
\caption{All feature requests for the screen \textit{Home Feed}, based on input from 40 activist social media users who experienced hate speech directed at them. The table shows importance ratings from Wave~1 and Wave~2, as well as the final ranking and the percentage of points from Wave~3. Requests are ordered by their Wave~3 rank.}
    \begin{tabular}{l l l c}
    \toprule
    \textbf{Feature Requests \textit{Home Feed}} & \textbf{Importance W1} & \textbf{Importance W2} & \textbf{Rank W3 (\% Points )}\\
    \midrule
    Differentiating Between Hate and Counter Speech & - &  \ding{52} 91.2\% \ding{56} 2.9\% & 1. (16.3\%)\\
    Customizing the Response to Detected Hate Speech &  \ding{52} 96.3\% \ding{56} 3.7\% & \ding{52} 94.1\% \ding{56} 0.0\%& 2. (14.0\%)\\
    Filtering Based on Words  &  \ding{52} 77.8\% \ding{56} 11.1\% & \ding{52} 97.1\% \ding{56} 2.9\% & 3.(11.5\%)\\
    Filtering Based on Users  & - & \ding{52} 88.2\% \ding{56} 2.9\%& 4. (9.7\%)\\
    Context-Sensitive Filtering & - &  \ding{52}  67.6\% \ding{56} 17.6\% & 5. (9.4\%)\\
    Filtering Based on Topics  &  \ding{52} 81.5\% \ding{56} 14.8\%&  \ding{52}  91.2\%\ding{56} 0.0\% & 6. (8.9\%)\\
    Filtering Based on Image Content  & - & \ding{52}  76.5\% \ding{56} 14.7\% & 7. (8.7\%)\\
    Filtering Based on Video Content  & - & \ding{52}  73.5\% \ding{56} 14.7\% & 8. (8.2\%)\\
    Filtering Based on Audio Content  & - & \ding{52}  67.6\% \ding{56} 14.7\% & 9. (6.9\%)\\
    Filtering Based on Examples  & - & \ding{52}  67.6\% \ding{56} 5.9\% & 10. (6.4\%)\\
    \bottomrule
\end{tabular}
\label{tab:HomeFeed}
\end{table}

\begin{table}
\Description{Table 13 summarizes all feature requests for the screen All DMs, based on input from 40 activist social media users targeted by hate speech. The table has four columns and eight rows. The first column, Feature Requests, lists the requested features: Customizing the Response to Detected Hate Speech, Differentiating Between Hate and Counter Speech, Filtering Based on Communication-Patterns, Filtering Based on Users, Filtering Based on Words, Filtering Based on Image Content, Filtering Links, and Filtering Based on Language.
The second column, Importance Wave 1, shows the percentage of participants who rated each feature as important, indicated by a check symbol, and the percentage who did not, indicated by a cross symbol. The third column, Importance Wave 2, uses the same format to show ratings from the second wave. The fourth column, Rank Wave 3, presents the final ranking of each feature in Wave 3 along with the percentage of total points it received.}
\caption{All feature requests for the screen \textit{All DMs}, based on input from 40 activist social media users targeted by hate speech. The table shows importance ratings from Wave~1 and Wave~2, as well as the final ranking and the percentage of points from Wave~3. Requests are ordered by their Wave~3 rank.}
    \begin{tabular}{l l l l}
    \toprule
    \textbf{Feature Requests \textit{All DMs}} & \textbf{Importance W1} & \textbf{Importance W2} & \textbf{Rank W3 (\% Points )}\\
    \midrule
    Customizing the Response to Detected Hate Speech & \ding{52} 96.9\% \ding{56} 0.0\% &  \ding{52} 97.1\% \ding{56} 0.0\% & 1. (14.9\%)\\
    Differentiating Between Hate and Counter Speech & - &  \ding{52} 91.2\% \ding{56} 2.9\%  & 2. (14.6\%)\\
    Filtering Based on Communication-Patterns & - &  \ding{52} 76.5\% \ding{56} 8.8\% & 3. (12.2\%)\\
    Filtering Based on Users & - &  \ding{52} 85.3\% \ding{56} 2.9\% & 4. (11.9\%)\\
    Filtering Based on Words & \ding{52} 78.1\% \ding{56} 9.4\% & \ding{52} 94.1\% \ding{56} 2.9\% & 5. (10.0\%)\\
    Filtering Based on Image Content & - & - & 6. (7.8\%)\\
    Filtering Links & - &  \ding{52} 73.5\% \ding{56} 23.5\% & 7. (6.5\%)\\
    Filtering Based on Language & - & - & 8. (5.4\%)\\
    \bottomrule
\end{tabular}
\label{tab:AllDMs}
\end{table}

\begin{table}
\Description{Table 14 presents the feature requests for the “Comments Specific Posts” screen. The table has four columns (Feature Request, Importance W1, Importance W2, and Rank W3 with percentage points) and seven rows of features. Features are ordered by importance based on their Wave 3 ranking. The table summarizes how participants rated each feature across Waves 1 and 2 and shows their final ranking in Wave 3.
The seven features are:
(1) Customizing the Response to Detected Hate Speech, which received the highest ranking; (2) Filtering Based on Words; (3) Filtering Based on Topics; (4) Filtering Based on Users; (5) Automatic Adjustment of Filter Strength; (6) Automatic Reference to the Community Guidelines; and (7) Filtering Based on Tags.
For each feature, the table reports the percentage of participants who rated it as important or not important in Waves 1 and 2, followed by its Wave 3 rank and percentage of total points. Overall, the top three features consistently received high importance ratings across waves, while later-ranked features were either added in later waves or received more mixed evaluations.}
\caption{All feature requests for the screen \textit{Comments Specific Posts}, based on input from 40 activist social media users targeted by hate speech. The table shows importance ratings from Wave~1 and Wave~2, as well as the final ranking and the percentage of points from Wave~3. Requests are ordered by their Wave~3 rank.}
    \begin{tabular}{l l l l}
    \toprule
    \textbf{Feature Requests \textit{Comments Specific Posts}} & \textbf{Importance W1} & \textbf{Importance W2} & \textbf{Rank W3 (\% Points )}\\
    \midrule
    Customizing the Response to Detected Hate Speech & \ding{52} 97.9\% \ding{56} 0.0\% &\ding{52} 94.3\% \ding{56} 0.0\%& 1. (20.8\%) \\ 
    Filtering Based on Words & \ding{52} 93.6\% \ding{56} 0.0\% & \ding{52} 94.3\% \ding{56} 2.9\%  & 2. (18.0\%)\\
    Filtering Based on Topics & \ding{52} 72.3\% \ding{56} 10.6\% & \ding{52} 82.9\% \ding{56} 2.9\%  & 3. (14.7\%)\\ 
    Filtering Based on Users & - &  \ding{52} 85.7\% \ding{56} 5.7\% & 4. (14.7\%)\\
    Automatic Adjustment of Filter Strength & - & \ding{52} 77.1\% \ding{56} 2.9\% & 5. (12.7\%)\\
    Automatic Reference to the Community Guidelines & - & - &6. (12.7\%) \\
    Filtering Based on Tags & - & \ding{52} 65.7\% \ding{56} 17.1\% &7. (7.3\%)\\
    \bottomrule
\end{tabular}
\label{tab:CommentsSpecificPosts}
\end{table}

\begin{table}
\Description{Table 15 presents the feature requests for the “DMs Specific Accounts” screen. The table has four columns (Feature Request, Importance W1, Importance W2, and Rank W3 with percentage points) and seven rows of features. Features are ordered by importance based on their Wave 3 ranking. The table summarizes how participants rated each feature in Waves 1 and 2 and shows their final ranking in Wave 3.
The seven features are: (1) Customizing the Response to Detected Hate Speech, which received the highest ranking; (2) Filtering Based on Words; (3) Follower-Based Blocking of DMs; (4) Filtering Based on Trust Ratings; (5) Filtering Based on Users; (6) Filtering Based on Topics; and (7) Context-Sensitive Filtering.
For each feature, the table reports the percentage of participants who rated it as important or not important in Waves 1 and 2, followed by its Wave 3 rank and percentage of total points. Overall, the top two features show strong and consistent support, while mid-ranked features display more mixed importance ratings, particularly those introduced in later waves.}
\caption{All feature requests for the screen \textit{DMs Specific Accounts}, based on input from 40 activist social media users targeted by hate speech. The table shows importance ratings from Wave~1 and Wave~2, as well as the final ranking and the percentage of points from Wave~3. Requests are ordered by their Wave~3 rank.}
    \begin{tabular}{l l l l}
    \toprule
    \textbf{Feature Requests \textit{DMs Specific Accounts}} & \textbf{Importance W1} & \textbf{Importance W2} & \textbf{Rank W3 (\% Points )}\\
    \midrule
    Customizing the Response to Detected Hate Speech & \ding{52} 94.7\% \ding{56} 0.0\% & \ding{52} 94.7\% \ding{56} 0.0\%  & 1. (23.3\%) \\ 
    Filtering Based on Words & \ding{52} 89.5\% \ding{56} 5.3\% & \ding{52} 91.7\% \ding{56} 2.8\%& 2. (16.0\%)\\ 
    Follower-Based Blocking of DMs & - & \ding{52} 68.6\% \ding{56} 8.6\% & 3. (15.1\%)\\
    Filtering Based on Trust Ratings & - & \ding{52} 76.5\% \ding{56} 14.7\%& 4. (14.3\%)\\ 
    Filtering Based on Users & - &  \ding{52} 88.9\% \ding{56} 2.8\% & 5. (14.3\%)\\
    Filtering Based on Topics & \ding{52} 65.8\% \ding{56} 15.8\% & \ding{52} 83.3\% \ding{56} 2.8\%& 6. (11.8\%)\\ 
    Context-Sensitive Filtering & - & - & 7. (5.3\%)\\
    \bottomrule
\end{tabular}
\label{tab:DMsSpAc}
\end{table}

\begin{table}
\Description{Table 16 presents the feature requests for the “Explore Tab” screen. The table has four columns (Feature Request, Importance W1, Importance W2, and Rank W3 with percentage points) and eight rows of features. Features are ordered by importance based on their Wave 3 ranking. The table summarizes how participants rated each feature in Waves 1 and 2 and shows their final ranking in Wave 3.
The eight features are:
(1) Customizing the Response to Detected Hate Speech, which received the highest ranking; (2) Filtering Based on Words; (3) Filtering Based on Topics; (4) Filtering Based on Image Content; (5) Filtering Based on Tone; (6) Filtering Based on Community Ratings; (7) Warnings for Content from Unknown Accounts; and (8) Filtering Based on Virality of Content.
For each feature, the table reports the proportion of participants who rated it as important or not important in Waves 1 and 2, followed by its Wave 3 rank and percentage of total points. Overall, the top three features show strong support across waves, while later-ranked features, particularly those introduced in Wave 2, received more varied importance ratings.}
\caption{All feature requests for the screen \textit{Explore Tab}, based on input from 40 activist social media users targeted by hate speech. The table shows importance ratings from Wave~1 and Wave~2, as well as the final ranking and the percentage of points from Wave~3. Requests are ordered by their Wave~3 rank.}
    \begin{tabular}{l l l l}
    \toprule
    \textbf{Feature Requests \textit{Explore Tab}} & \textbf{Importance W1} & \textbf{Importance W2} & \textbf{Rank W3 (\% Points )}\\
    \midrule
    Customizing the Response to Detected Hate Speech & \ding{52} 86.0\% \ding{56} 4.7\% &\ding{52} 91.2\% \ding{56} 2.9\% & 1. (20.2\%)\\
    Filtering Based on Words & \ding{52} 72.1\% \ding{56} 14.0\% & \ding{52} 91.2\% \ding{56} 2.9\% & 2. (14.7\%)\\
    Filtering Based on Topics & \ding{52} 74.4\% \ding{56} 11.6\% & \ding{52} 94.1\% \ding{56} 0.0\% & 3. (14.3\%)\\
    Filtering Based on Image Content & - & \ding{52} 79.4\% \ding{56} 11.8\% & 4. (14.3\%)\\
    Filtering Based on Tone & - & \ding{52} 64.7\% \ding{56} 14.7\% & 5. (12.1\%)\\
    Filtering Based on Community Ratings &- & \ding{52} 67.6\% \ding{56} 11.8\% & 6. (9.2\%)\\
    Warnings for Content from Unknown Accounts & - & - & 7. (9.2\%)\\
    Filtering Based on Virality of Content &- & - & 8. (5.9\%)\\
    \bottomrule
\end{tabular}

\label{tab:ExploreTab}
\end{table}

\end{document}